\documentclass[sort&compress,3p,12pt]{elsarticle}
\journal{Annals of Physics}

 \usepackage{amsmath,amssymb,bm,euscript,eufrak,mathrsfs,bbm}
 \usepackage{graphicx,boxedminipage}
 \usepackage[format=plain]{caption}
 \usepackage{floatflt}
 \usepackage[hypertex]{hyperref}
 \usepackage{xcolor,fancybox}
\newcommand{\bra}[1]{\ensuremath{\bm{\langle}#1\bm{|}}}
\newcommand{\ket}[1]{\ensuremath{\bm{|}#1\bm{\rangle}}}
\newcommand{\Tr}{\ensuremath\mathrm{Tr\,}}
\newcommand{\Sp}{\ensuremath\mathrm{Sp\,}}

\newcommand{\av}[1]{\ensuremath{\langle#1\rangle}}
\newcommand{\Av}[1]{\ensuremath{\big<#1\big>}}
\newcommand{\AV}[1]{\ensuremath{\Big<#1\Big>}}
\newcommand{\openone}{\mathbbmss{1}}
\allowdisplaybreaks

\begin{document}

\begin{frontmatter}

\title{Dual nature of localization in guiding systems with randomly corrugated boundaries: Anderson-type versus entropic}

\author{Yu.\,V. Tarasov\corref{cor1}}
\ead{yutarasov@ire.kharkov.ua}
\cortext[cor1]{Corresponding author}

\author{L.\,D. Shostenko\corref{cor0}}

\address{Institute for Radiophysics and Electronics, NAS of Ukraine,\\
         12 Proskura Str., Kharkiv 61085, Ukraine}

\begin{abstract}
   A unified theory for the conductance of an infinitely long multimode quantum wire whose finite segment has randomly rough lateral boundaries is developed. It enables one to rigorously take account of all feasible mechanisms of wave scattering, both related to boundary roughness and to contacts between the wire rough section and the perfect leads within the same technical frameworks. The rough part of the conducting wire is shown to act as a mode-specific randomly modulated effective potential barrier whose height is governed essentially by the asperity slope. The mean height of the barrier, which is proportional to the average slope squared, specifies the number of conducting channels. Under relatively small asperity amplitude this number can take on arbitrary small, up to zero, values if the asperities are sufficiently sharp. The consecutive channel cut-off that arises when the asperity sharpness increases can be regarded as a kind of localization, which is not related to the disorder \emph{per se} but rather is of entropic or (equivalently) geometric origin. The fluctuating part of the effective barrier results in two fundamentally different types of guided wave scattering, viz., inter- and intramode scattering. The intermode scattering is shown to be for the most part very strong except in the cases of (\emph{a})~extremely smooth asperities, (\emph{b})~excessively small length of the corrugated segment, and (\emph{c})~the asperities sharp enough for only one conducting channel to remain in the wire. Under strong intermode scattering, a~new set of conducting channels develops in the corrugated waveguide, which have the form of asymptotically decoupled extended modes subject to individual solely intramode random potentials. In view of this fact, two transport regimes only are realizable in randomly corrugated multimode waveguides, specifically, the ballistic and the localized regime, the latter characteristic of one-dimensional random systems. Two kinds of localization are thus shown to coexist in waveguide-like systems with randomly corrugated boundaries, specifically, the entropic localization and the one-dimensional Anderson (disorder-driven) localization. If the particular mode propagates across the rough segment ballistically, the Fabry--P\'erot-type oscillations should be observed in the conductance, which are suppressed for the mode transferred in the Anderson-localized regime.
\end{abstract}

\date{\today}

\begin{keyword}
  wave localization \sep
  multichannel quantum waveguide \sep
  surface disorder \sep
  entropic potential \sep
  gradient scattering mechanism
\end{keyword}

\end{frontmatter}

\section{Introduction}

The multiple scattering of quantum and classical waves in disordered systems is at the heart of many interesting physical phenomena. Among them the best known and thoroughly studied are the Anderson localization of current carriers and waves of different types, quantum corrections to conductivity (weak localization), metal-insulator transitions in solid state systems, and the coherent retro-reflection of waves (the backscattering enhancement) \cite{LifGredPast88,Sheng06,AkkermMontamb07}. Recently, Bose-condensed systems with numerous manifestations of state localization, which are in one way or another related to random potentials, have also become the subject of active research (see Refs.~\cite{Lewenstein07,Modugno10,PalenciaLewenstein10,Shapiro12} and references therein).

In recent years, however, there has been an increasing tendency for obtaining the experimental results that either were not in good conformity with conventional theories of disorder-induced localization or even substantially inconsistent with them (see, e.\,g., Ref.~\cite{NguyenKim12} and references therein). The observed inconsistencies may have different origins, to which one can attribute, for instance, uncommon refractive properties of propagation media \cite{Asatryan07}, specific correlation properties of scattering potentials \cite{IzrKrokhMak12}, the availability of absorption or amplification sources \cite{ScalesCarr07}, and other possible factors. The non-trivial topology of propagation medium may also play a major role. The latter factor manifests itself in diverse phenomena such as Aharonov-Bohm and quantum Hall effects \cite{AharonovBohm59,Halperin82}, non-homogeneous concentration of energy in randomly filled cavities \cite{Sapienza10}, unusual properties of current carrier transport in wires both with isolated bends (in Ref.~\cite{Timp88}, the ``bent resistance'' has been detected experimentally) and in ballistic conductors with randomly distorted boundaries \cite{MakTar98,MakTar01}.

The random nature of systems of waveguide configuration (quantum conductors and classical waveguides) and the consequent scattering results in waveguide state attenuation commonly characterized by the \emph{extinction} length \cite{BassFuks79}, and also in spacial localization of those states, which has the interference nature and is specified by the \emph{localization} length. Strictly speaking, the theory of such a localization is valid for systems that are infinitely large in the direction of waveguide axis. In the present study we concentrate on the problem of electron (as well as classical wave) transport across a \emph{finite} (in the direction of current) disordered systems whose random nature is ensured by random roughness of their side boundaries.

Any deformation of the waveguide system boundaries is known to result in the appearance of the so-called \emph{entropic barriers} \cite{Zwanzig92}, whose notion is closely associated with the definition of \emph{quantum channels} in the information transfer theory \cite{Holevo12}. In contrast to energetic barriers, which are originally present in the Hamiltonian, the entropic barriers are not of the potential relief nature. A possible way to take them into account in practice is to introduce the emulating potential terms into the Hamiltonian, whose effect would be the same as in strictly considering the scattering in systems with destroyed boundaries \cite{Tesanovic86}. Yet another approach, which is more correct mathematically, was applied in Refs.~\cite{MeyerStep95,BratkRashk96}, where boundary irregularities were converted into the scattering potentials of Schr\"odinger Hamiltonian by the appropriately chosen ``smoothing'' coordinate transformation.\footnote{The idea of the ``smoothing'' coordinate transformation goes back to Migdal's work on the deformed nuclei theory, see Ref.~\cite{Migdal77}. As to the application of such transformations to wave scattering from rough surfaces the interested reader is referred to papers by Fedders \cite{Fedders68} and Celli \emph{et al.} \cite{Celli75}.}
In Refs.~\cite{MakTar98,MakTar01}, one more method for transforming the entropic barriers into the energetic ones was proposed, viz., the transition in Schr\"odinger equation to the representation of local (i.\,e., lengthwise-coordinate-dependent) transverse quantization modes.

In the two latter of the above-listed approaches a serious difficulty arose, which was not overcome for a certain period of time. The challenge was that if the scattering caused by effective potentials is taken into account, all the modes turned out to be noticeably intermixed, and their strict separation was thought of as impossible in the general case. To minimize the mode entanglement, in Refs.~\cite{MakTar98,MakTar01,MeyerStep95,BratkRashk96}, as well as in most of other works on the subject, substantial limitations were imposed on the boundary profile, of which the most fundamental was the requirement for the asperities to be sufficiently smooth.

Here it should be noted that until recently the numerous problems pertinent to wave (both classic and quantum) scattering at rough interfaces were amenable to the solution under quite strong limitations only, of which the most vital is the condition for inhomogeneity-induced scattering to be considered as weak. The most popular approximations that have resulted in the majority of well-known analytic results are the small-amplitude approximation, the Kirchhoff (tangent plane) approximation, the Born approximation for Lippmann-Schwinger equation, and Voronovich's small-slope approximation, all reviewed in Refs.~\cite{Maradudin07,Simonsen10}. Within all these approaches the asperities are regarded to be sufficiently smooth, in that the mean tangent of their slope relative to the unperturbed boundary is small as compared with unity. If this requirement is violated, the results could be obtained by numerical methods only.

The limitation to smooth roughness in the studies of quantum and classical wave surface scattering suffices to build up consistent perturbation theories. Yet, for the scattering to be regarded as weak the requirement for asperities to be smooth is not mandatory, really. Within the quantum mechanics approach the scattering probability is known to be specified by the integral rather than local quantities, namely, by the potential matrix elements. Therefore, by formulating the scattering problem in terms of effective potentials we are allowed, in principle, to go beyond the restriction by smooth asperities.

For analytical calculations the going beyond the smooth asperity limit was initially performed in Ref.~\cite{GanTarShost11}, where the spectral problem for cylindrical cavity resonator with randomly rough side boundary was solved with no limitations on the asperity sharpness degree. The further advancement was made in Ref.~\cite{GorTarShost13}, where, contrary to Ref.~\cite{GanTarShost11}, the essentially open system was investigated, specifically, the multimode waveguide whose side boundaries on the finite length interval are subject to periodical corrugation. In that paper, in view of non-random nature of the system under consideration, the problems somehow related to Anderson localization were not dealt with. Yet, some kind of localization was revealed in Ref.~\cite{GorTarShost13}, though its nature is notably different from the localization of Anderson origin. The essence of localization disclosed in Ref.~\cite{GorTarShost13} is that the modes extended in the infinite terminals of the wave-guiding system on strengthening their scattering within the modulated segment are sequentially transformed to the evanescent-type modes, thus being excluded from the set of those modes the energy is transported through. As the boundary corrugation profile is sharpened, the wave conductance of such systems is reduced in modulated-step fashion, where on each of the plateaus the conductance is subject to oscillations of Fabry-P\'erot (FP) origin. This demonstrates the extended mode partial trapping within the modulated segment that acts thereby as an open cavity resonator.

In the present paper we examine electron transport through the finite segment of an infinitely long two-dimensional (2D) conductor which does not contain ``bulk'' imperfections but rather has randomly rough lateral boundaries on that segment. The average width of the conducting strip is regarded as arbitrary, the current carriers are thought to be in the ground state. Similar problem was solved previously in Refs.~\cite{MakTar98,MakTar01}, where the one-mode metal strip was considered, which, to a certain accuracy, can be regarded as a strictly 1D disordered quantum system. In compliance with the general theory of such systems, all the electron states in that strip were proven to be exponentially localized at an arbitrary level of boundary roughness. The localization length was shown to be strongly dependent on the specific physical mechanism of electron-boundary scattering. Two fundamentally different scattering mechanisms were discriminated in Refs.~\cite{MakTar98,MakTar01}. For one of them, the mean amplitude of boundary asperities was shown to serve as the governing parameter whereas the other was proven to be governed mainly by the asperity rms slope. Both of the mechanisms contribute to the scattering amplitude additively, hence they were called the ``height'' and the ``slope'' scattering mechanisms, respectively. In the subsequent papers devoted to the subject (see Ref.~\cite{Izr03}), the slope mechanism was more reasonably named the ``gradient'' one, so this particular term will further be used in the present paper.

A natural extension of studies performed in Refs.~\cite{MakTar98,MakTar01} is the analysis of electron and wave transport in \emph{multimode} waveguide-type systems, both quantum and classical. The systems of this type, conventionally referred to as \emph{quasi-one-dimensional} (Q1D) systems, were deeply examined by a number of authors. The best known results on their localization properties were obtained in Refs.~\cite{Dorokhov84,Dorokhov88,MelPerKum88}, whereupon in the scientific community the belief was permanently established that in Q1D conductors all electron states are localized at an arbitrary level of the disorder. Yet later on, in Ref.~\cite{Tar00}, it was shown that in the general case this statement is not quite correct. Specifically, if the disorder is weak enough and all the scattering channels (waveguide modes) are non-equivalent, the exponential Anderson localization does not occur because of the eigenstate dephasing that is caused by the chaotic intermode scattering.

It should be noted that in Refs.~\cite{Dorokhov84,Dorokhov88,MelPerKum88,Tar00} the disorder of bulk origin only was taken into consideration. Meanwhile, according to some researchers (see, e.\,g., Refs.~\cite{NikolicMacKinnon94,SanchFreilYurkMarad98,SanchFreilYurkMarad99}) there should be a fundamental difference between scattering resulting from bulk and surface disorder in waveguide-like systems. The disorder of the latter type brings about quite unexpected results, among which is the hierarchy of localization lengths pertinent to different channels \cite{SanchFreilYurkMarad98,SanchFreilYurkMarad99,RendMakIzr11}. The multiplicity of localization lengths predicted for surface-disordered waveguides contrasts significantly with the existence of unique localization length typical for systems disordered in the bulk. The existence of such a hierarchy seems to be a non-trivial fact, because the irregularity in the waveguide, irrespective of its bulk or surface nature, gives rise to the intermode scattering and, accordingly, to the mode entanglement, thereby leading to violation of mode independence and, hence, to that of their uniqueness.

In Refs.~\cite{SanchFreilYurkMarad98,SanchFreilYurkMarad99}, the conclusion about the localization length hierarchy was made basically from numerical analysis. A similar result of Ref.~\cite{RendMakIzr11} was obtained by invoking some analytics. However, the calculation technique applied in the latter research offered the potential to manipulate with relatively small number of propagating modes. Specifically, two modes only were taken into account therein. The goal of the present paper is to examine waveguide-like systems with arbitrary number of propagating modes. The capability for thorough analysis of such systems is provided by the specific operator technique of mode separation previously developed in Ref.~\cite{Tar00} for conducting wire-like systems subject to arbitrary static potentials. Using this technique, any multidimensional transport problem can be reduced to solution of a countable set of the same-type 1D problems, in the general case non-Hermitian. Such a reduction may be regarded as a particular ``separation of variables'' for systems that were for long believed to be non-integrable. Recent applications of this technique to the problem of wave transport through a waveguide segment with periodically modulated side walls \cite{GorTarShost13} made it possible to ascertain that, firstly, in systems with corrugated boundaries the prevailing mechanism of corrugation-induced wave scattering is the gradient mechanism, the intensity of which is determined by the asperity sharpness rather than by their amplitude. Secondly, the technique applied in Ref.~\cite{GorTarShost13} has allowed us to describe purely analytically the scattering caused by asperities of any sharpness, not by smooth asperities only. This, in turn, permitted us to discover the peculiar sharpness-induced mode cut-off effect, whose essence consists in that all extended modes of the side-modulated waveguide are sequentially transformed into the modes of evanescent type on sharpening the modulation profile, thus becoming localized right at the entrance to the waveguide rippled section. At some critical value of the sharpness parameter the waveguide turns into the beyond-cutoff waveguide, even though the amplitude of corrugation is rather small and insufficient for that purpose.

The system examined in Ref.~\cite{GorTarShost13} was absolutely deterministic. Thus, the mode localization in the form of their transformation from extended to evanescent type has nothing in common with the conventional Anderson localization. In the present study we intend to examine conjointly the localization of both types, geometric and Anderson, to elucidate their role in electron transmission through quantum waveguide finite segments with boundaries corrugated randomly.

\section{Statement of the problem}

\begin{figure}[h]
  \setcaptionmargin{.5in}%
  \centering
  \scalebox{.8}[.8]{\includegraphics{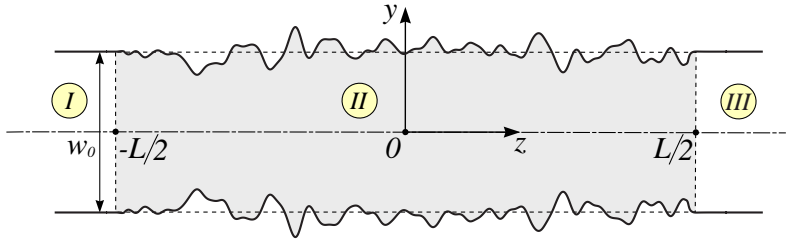}}
  \caption{Two-dimensional quantum waveguide with symmetric randomly corrugated boundaries on segment $\mathbb{L}: z\in(-L/2,L/2)$ (shaded).
  \hfill\label{fig1}}
\end{figure}

Consider infinitely long 2D conducting strip (quantum waveguide) with straight parallel walls, whose side boundaries on the finite segment of length $L$ are subjected to corrugation of random nature (see Fig.~\ref{fig1}). In the general case the fluctuations of opposite walls are described by different random functions, either cross-correlated
or not. But in this study we will discuss the instance of boundaries corrugated \emph{symmetrically} with respect to the strip central axis. In Ref.~\cite{MakTar01} it was shown that such a model of the rough quantum waveguide takes correctly into account (at least qualitatively) all surface scattering mechanisms, both the amplitude and the gradient mechanism.

If the deviation of one of the conductor boundaries from the straight line is specified by function $\xi(z)$, then its local width (along axis $z$) equals $w(z)=w_0+2\xi(z)\theta(L/2-|z|)$. By $w_0$ we denote the width of the non-corrugated conductor leads (sections \textit{I} and \textit{III} in Fig.~\ref{fig1}). We specify random process $\xi(z)$ by correlation equalities
\begin{subequations}\label{Xi-corr_equalities}
\begin{align}\label{Av_xi=0}
  \Av{\xi(z)} &= 0\ ,\\
\label{Xi_bin_corr}
  \Av{\xi(z)\xi(z')} &= \sigma^2W(z-z')\ ,
\end{align}
\end{subequations}
where  $\sigma$ is the boundary rms deviation, $W(z)$ is the function with unit maximum at $z=0$, which decreases to negligibly small values at a distance $|\Delta z|\sim r_c$ (correlation radius) from this point. The mean amplitude of the corrugation is assumed to be small as compared to the waveguide average width,
\begin{equation}\label{Small_ampl}
  \sigma\ll w_0\ .
\end{equation}
At the same time, the sharpness of the corrugation profile, which is specified by ratio $\sigma/r_c$, will be regarded as arbitrary.\footnote{
The natural measure of the asperity sharpness degree is the ratio of their rms height to the mean longitudinal dimension. Although in the general case the correlation radius may differ significantly from the dimension of boundary cambers and dips (e.\,g., for periodic corrugation the correlation length is at all infinite), in this paper we rely on the results of Ref.~\cite{MaradMichel90} according to which the correlation length for \emph{random} asperities of the general type coincides, on the order of magnitude, with their longitudinal extent.
}

When calculating the conducting properties of the system depicted in Fig.~\ref{fig1} we proceed from Landauer's approach \cite{HaeringenLenstra90},  which in the statical limit is known to be fully equivalent to Kubo's linear response theory \cite{FisherLee81}. According to the Landauer approach, the average conductance of the finite-length conducting system taken as a multimode quantum waveguide equals the product of conductance quantum $g_0=e^2/2\pi\hbar$ and the complete transmittance of the waveguide, $\Tr(\hat{t}\hat{t}^{\dag})$, where $\hat{t}=\big\{t_{mn}\big\}$ is the matrix composed of transmission coefficients from channel $n$ to channel~$m$,
\begin{equation}\label{Cond->t_mn}
  g(L)=g_0\sum_{m,n}\left|t_{mn}(L)\right|^2\ .
\end{equation}
As is shown in Ref.~\cite{StoneSzafer88} (see also Ref.~\cite{Zirnbauer92}), the transmission coefficients entering into equality~\eqref{Cond->t_mn} can be expressed in terms of matrix elements of the \emph{end-to-end} Green function, of which the ``source'' point is located at one end of the conductor whereas the ``receiver'' is at the other end.

We will seek for Green function as the solution to equation
\begin{equation}\label{Schrodinger_Green}
  (\Delta+k^2)G(\mathbf{r},\mathbf{r}')=\delta(\mathbf{r}-\mathbf{r}')
\end{equation}
(the units with $\hbar=2m=1$ are used), where $\Delta$ is the Laplace operator and $k^2=\varepsilon_F$ is the Fermi energy. As a first step to solve this equation, we proceed to the representation of waveguide modes by Fourier-transforming equation~\eqref{Schrodinger_Green} over transverse coordinate $y$ using locally (at each $z$) complete set of eigenfunctions
\begin{align}\label{Whole_set}
 & \ket{y,z;n} =\left[\frac{2}{w(z)}\right]^{1/2}
  \sin\left\{\left[\frac{y}{w(z)}+\frac{1}{2}\right]\pi n\right\}
  \qquad (n=1,2,3,...)\ ,
\end{align}
which obey orthonormality condition $\bm{\langle}y,z;n\bm{|}y,z;n'\bm{\rangle}=\delta_{nn'}$ with scalar product defined as the integral over $y$ variable that runs the waveguide local cross-section, specifically, the interval $-w(z)/2<y<w(z)/2$. In such a way the Green function can be presented in the form of double Fourier series,
\begin{equation}\label{Green_ModeRep}
  G(\mathbf{r},\mathbf{r}') = \sum_{n,n'=1}^{\infty}
  \bra{y,z;n}G_{nn'}(z,z')\ket{y',z';n'} \ ,
\end{equation}
with coefficients (Green function mode components $G_{nn'}$) that obey the following set of coupled equations,
\begin{equation}\label{G_mode-eqn}
  \left[\frac{\partial^2}{\partial z^2}+
  k^2_n-V_n(z)\right]G_{nn'}(z,z')
  -\sum_{m=1\atop (m\neq n)}^{\infty}\hat{U}_{nm}(z)
  G_{mn'}(z,z')=
  \delta_{nn'}\delta(z-z') \ .
\end{equation}
In equation~\eqref{G_mode-eqn}, $k_n^2=k^2-\left(\pi n/w_0\right)^2$ is the longitudinal wavenumber of $n$-th waveguide mode in the non-corrugated leads, $V_n(z)$ and $\hat{U}_{nm}(z)$ are the ``effective potentials'' which arise in the wave equation on account of boundary roughness and which we regard as the origin of the intramode ($V_n$) and the intermode ($\hat{U}_{nm}$) scattering for the outward waveguide modes. The structure of these potentials is as follows,
\begin{subequations}\label{Potentials_VnUnm}
 \begin{align} \label{Vn}
  & V_n(z)=\theta(L/2-|z|)\left\{\pi^2n^2\left[\frac{1}{w^2(z)}-\frac{1}{w_0^2}\right]+
   \left(1+\frac{\pi^2n^2}{3}\right)
   \left[\frac{w'(z)}{2w(z)}\right]^2\right\}\ ,\\
  \label{Unm}
  & \hat{U}_{nm}(z)=\theta(L/2-|z|)2B_{nm}\left\{D_{nm}\left[\frac{w'(z)}{w(z)}\right]^2-
   \frac{1}{w(z)}\left[w'(z)\frac{\partial}{\partial z}+\frac{\partial}{\partial z}w'(z)\right]\right\}\ ,
 \end{align}
\end{subequations}
numerical coefficients in \eqref{Unm} are given as
\begin{subequations}\label{BnmDnm}
 \begin{align}\label{Bnm}
   & B_{nm}=\frac{nm}{n^2-m^2}\cos^2\left[\frac{\pi}{2}(n-m)\right]\ ,\\
 \label{Dnm}
   & D_{nm}=\frac{3n^2+m^2}{n^2-m^2}\ .
 \end{align}
\end{subequations}

The set of equations \eqref{G_mode-eqn} should be supplemented with appropriate boundary conditions (BCs) on axis $z$. Since the waveguide by definition is an open system, as the BCs at ${z\to\pm\infty}$ we apply the so-called ``radiation'' conditions. To formulate them in analytical form for the nonuniform waveguide system of arbitrary cross-section is a challenging task in the general case. However, in view of potentials \eqref{Potentials_VnUnm} being identically equal to zero in the regions where $|z|>L/2$, the Green function mode content in these regions reduces to diagonal components only, so the legible BCs may be \textit{a~priori} stated exclusively for these components. In the next section it will be shown that this would suffice to determine the entire Green function since all the nondiagonal components of its mode matrix are linearly expressed through the diagonal components, for which a countable set of uncoupled equations will be obtained.

\section{Mode separation in the nonuniform quantum waveguide}
\label{ModeSep}

The set of equations \eqref{G_mode-eqn} can be solved with respect to Green function mode components through the operator procedure applicable for potentials $V_n(z)$ and $\hat{U}_{nm}(z)$ of quite arbitrary form \cite{Tar00}. Although this method was originally developed for open waveguide-like systems with potentials of volume nature, it was also expanded to closed systems in subsequent works, with the disorder available both in the bulk \cite{GanErTar07} and on the surface \cite{GanTarShost11}.

Briefly, with regard to waveguide-like systems the method reduces to the following set of mathematical steps. At the first stage, by putting $n\neq n'$ in Eq.~\eqref{G_mode-eqn}, the set of equations with all $n$'s should be solved for nondiagonal elements of mode matrix $\|G_{mn}\|$ to express them, by linear operation, in terms of the diagonal elements whose mode indices are coincident with right-hand (column) index of the sought-for nondiagonal element,
\begin{equation}\label{G_mn->G_nn}
  G_{mn}(z,z')=\int_{\mathbb{L}}\mathrm{d}z_1\,
  \mathsf{K}_{mn}(z,z_1)G_{nn}(z_1,z')
  \qquad(m\neq n) \ .
\end{equation}
In Eq.~\eqref{G_mn->G_nn}, the integration is done over interval $\mathbb{L}\!\!: z\in(-L/2,L/2)$, where potentials~\eqref{Potentials_VnUnm} are not identically equal to zero. The kernel of the integral operator in \eqref{G_mn->G_nn} can be found from the equation of Lippmann-Schwinger type,
\begin{equation}\label{K_numu}
  \mathsf K_{mn}(z,z')={\mathsf R}_{mn}(z,z')
  +\sum_{k\neq n}
  \int_{\mathbb{L}}dz_1\,\mathsf{R}_{mk}(z,z_1)
  \mathsf{K}_{kn}(z_1,z') \ ,
\end{equation}
where
\begin{equation}\label{kernR}
  \mathsf{R}_{mn}(z,z')=G^{(V)}_{m}(z,z')U_{mn}(z')
\end{equation}
is the matrix element of some operator $\hat{\mathsf{R}}$ defined in coordinate-mode space $\mathsf{M}=\{z,m\}$ that includes coordinate axis $z$ and the total number of mode indices.

Function $G^{(V)}_{m}(z,z')$ in definition \eqref{kernR} (in what follows we will refer to it as the \emph{trial} mode Green function) is the solution of equation
\begin{equation}\label{G(V)_m-eq}
  \left[\frac{\partial^2}{\partial z^2}+
  k^2_m-V_m(z)\right]G^{(V)}_m(z,z') =\delta(z-z')\ ,
\end{equation}
which differs from Eq.~\eqref{G_mode-eqn} by the absence of intermode potentials. By solving~\eqref{G(V)_m-eq} and considering operator $\hat{\mathsf{R}}$ matrix elements as given functions we can substitute nondiagonal elements of Green matrix $\|G_{mn}\|$ into Eq.~\eqref{G_mode-eqn}, thus arriving at the set of same-type closed equations for the diagonal elements of this matrix,
\begin{equation}\label{GDIAG-FIN}
  \left[\frac{\partial^2}{\partial z^2}+
  k^2_{n}-V_{n}(z)-\hat{\mathcal{T}}_{n}\right]
  G_{nn}(z,z')=\delta(z-z')\qquad\qquad (\text{for}\ \forall n) \ .
\end{equation}
Here, the additional potential $\hat{\mathcal{T}}_{n}$, if compared with equation~\eqref{G(V)_m-eq} for the trial Green function, has appeared. The operator expression for this potential has the form
\begin{equation}\label{T-oper}
  \hat{\mathcal{T}}_{n}=\bm{P}_{n}\hat{\mathcal U}
  (\openone-\hat{\mathsf R})^{-1}\hat{\mathsf R}\bm{P}_{n}\ .
\end{equation}
Here $\hat{\mathcal U}$ is the operator intermode potential specified in $\mathsf{M}$ by matrix elements
\begin{equation}\label{U-matr}
  \bra{z,n}\hat{\mathcal U}\ket{z',m} =
  {U}_{nm}(z)\delta(z-z')\ ,
\end{equation}
$\bm{P}_{n}$ is the projection operator whose effect reduces to the assignment of given value $n$ to
the nearest mode index of any operator standing next to it, no matter whether it is to the left or to
the right. The appearance of such operators is due to the fact that in view of the limitation of sums in formulas~\eqref{G_mode-eqn} and \eqref{K_numu} by the indices nonequal to $n$, the operators that stand between two projectors in Eq.~\eqref{T-oper} have the effect not in the whole space $\mathsf{M}$ but rather in its reduced version (the subspace ${\mathsf{\overline{M}}_{n}}\in\mathsf{M}$) which includes the coordinate axis $z$ and the set of all mode indices other than the separate index $n$. The role of the projection operators in Eq.~\eqref{T-oper} is to reduce the effect of this operator on the function located to the right of it (in our particular case, on the Green function $G_{nn}(z,z')$) solely to integration over~$z$, without summation over mode indices.

From the functional structure of potential  \eqref{T-oper} it may be inferred that in equation \eqref{GDIAG-FIN} for the diagonal propagator of a given transverse mode this potential virtually allows for the \emph{intermode} scattering. In contrast to potential $V_n(z)$, the potential $\hat{\mathcal{T}}_{n}$, in the general case, acts as the nonlocal (in $z$) operator whose characteristic scale is determined by the spatial extent of the trial Green functions and the primordial intermode potentials \eqref{Unm}, all entering into the \emph{T}-potential through the \emph{mode-mixing operator} $\hat{\mathsf{R}}$.

Now we give a brief comment on the boundary conditions to equation \eqref{GDIAG-FIN}. The $n$-th waveguide mode, whose dynamics is governed by this equation, proves to be effectively isolated from other modes, which enter into \eqref{GDIAG-FIN} only implicitly, as the intermediate scattering states ``hidden'' inside the \textit{T}-potential. If Eq.~\eqref{GDIAG-FIN} is rewritten as the integral Dyson equation, where by way of the unperturbed Green function the trial function $G^{(V)}_m(z,z')$ stands, it will be made clear that the exact Green function obeys the same (in form) BC as the trial function does, if the latter conditions are stated in the linear and homogeneous form. We will present the explicit BCs for the trial Green function in the next section.

Thus, taking into account relationship \eqref{G_mn->G_nn} between diagonal and nondiagonal elements of the Green matrix, which may be given as operator equality
\begin{equation}\label{Gmn->Gnn}
  \hat{G}_{mn}=\bm{P}_m(\openone-\hat{\mathsf R})^{-1}\hat{\mathsf R}\bm{P}_n \hat{G}_{nn}
\end{equation}
with operators $\hat{G}_{mn}$ and $\hat{G}_{nn}$ being thought of as matrices in the space of $z$ variable only, we can assert that the solution to the entire set of uncoupled equations \eqref{GDIAG-FIN} fully determines the sought-for Green function of the infinitely long conducting strip that contains the corrugated finite-length segment. The derivation of equation \eqref{GDIAG-FIN} and the potentials entering into it is, in fact, equivalent to separation of variables in the initial 2D equation \eqref{Schrodinger_Green} providing that the solution to Eq.~\eqref{G(V)_m-eq} is found independently. Such a two-stage approach largely facilitates the fulfillment of the initial task since a number of highly advanced mathematical techniques have been developed by now to solve strictly 1D equations.

Meanwhile, it should be admitted that the additional (to the original $V_n(z)$) effective intramode potential with quite nontrivial functional structure has appeared in the master equation at the expense of the achieved simplification of the initially stated problem. The analogous potential is known in the quantum scattering theory, being referred to as \emph{T}-matrix \cite{T75,N68}. However, the conventionally defined \emph{T}-matrix is known to be an extremely singular mathematical object, which is normally calculated in the lowest order of perturbation theory. In our calculation technique, the \emph{T}-potential in equation~\eqref{GDIAG-FIN} has no singularities. In Ref.~\cite{Tar00} it was shown that if the effective potentials in Eq.~\eqref{G_mode-eqn} are random functions, which break completely the spatial symmetries of the system under consideration, the regularity of potential \eqref{T-oper} is ensured by splitting the entire set of effective potentials into the subsets of intramode and intermode potentials. But even though the system would be subjected to regular but not to random perturbations, its openness and, therefore, the non-Hermitian property, will be sufficient for the inverse operator in \eqref{T-oper} to be singularity-free. This is evidenced in fundamental papers by Feshbach on nuclear reaction theory \cite{Feshbach58,Feshbach62} (see also Refs.~\cite{Feinberg09,Feinberg11}, where this issue is discussed in detail with regard to one-dimensional systems subject to final-support external potentials). Nevertheless, in order to ensure an additional protection against the degeneracy property of the full wave operator in \eqref{G_mode-eqn} one can supply the ``energy'' $k^2$ with the extra imaginary term that allows for infinitesimal dissipation, thus making the inverse operator in \eqref{T-oper} properly defined.

\section{The trial Green function evaluation}
\label{Trial_Green_func}

In the course of solution of basic equation \eqref{Schrodinger_Green} we have split the set of effective potentials arisen in going to mode representation into the subsets of intramode and intermode potentials (see Eqs.~\eqref{Potentials_VnUnm}). The potentials of the former type determine, through equation \eqref{G(V)_m-eq}, the set of \emph{trial} Green functions, whose mode index is preserved during the scattering. Allowing for this fact we will refer to the scattering induced by intramode potentials as the coherent scattering. Conversely, potentials \eqref{Unm} determine the scattering between the modes with different transverse energies, so we will refer to this type of scattering as the noncoherent one. In this section, we will analyze the coherent scattering only through calculation of trial propagator $G^{(V)}_n(z,z')$. This propagator, as will be seen below, plays an extremely important role in our study, so we will give a detailed account of its calculation.

Since all the transverse modes of the considered non-uniform system are effectively disentangled, the condition for the conductor openness in the direction of current flow can be formulated for each of the modes separately. At the conductor leads, the $n$'th mode longitudinal wavenumber equals $k_n$, so the openness condition for the system ``conductor $+$ leads'' yields equalities
\begin{equation}\label{BC_trial_GF}
  \left.\left(\frac{d}{dz}\mp ik_n\right)G^{(V)}_n(z,z')\right|_{z\to\pm\infty}=0
  \qquad(\text{for $\forall n$'s})\ .
\end{equation}
In such a statement the problem of determining function $G^{(V)}_n(z,z')$ from equation~\eqref{G(V)_m-eq} belongs to the class of open-type boundary-value (BV) problems, whose exact solution when the dimension is larger than one is difficult to obtain even for potentials of relatively simple form. The advantage of solving the exactly 1D problems is that each of them, being initially stated as a BV problem, can always be reduced to solving the problems of causal type. This is critical for performing the configurational averaging, since there is a pressing need for multiple scattering to be taken into account for one-dimensional disordered systems.

The transition from BV problem stated in the form of Eqs.~\eqref{G(V)_m-eq} and \eqref{BC_trial_GF} to corresponding Cauchy problems can be done through the following representation of 1D Green function (see, e.\,g., Ref.~\cite{CourantHilb53}),
\begin{equation}\label{Green-Cochi}
  G_n^{(V)}(z,z')=\mathcal{W}^{-1}_n(z')
  \big[ \psi_+(z|n)\psi_-(z'|n)\theta(z-z') +
  \psi_+(z'|n)\psi_-(z|n)\theta(z'-z) \big] \ .
\end{equation}
Here, functions $\psi_{\pm}(z|n)$ are the solutions of two auxiliary problems for homogeneous equation \eqref{G(V)_m-eq}, whose boundary conditions are specified on just one ("plus" or "minus", respectively) end of the interval where the Green function is being sought for; $\mathcal{W}_n(z)$ is the Wronskian of those solutions. By representing potential \eqref{Vn} as a sum of the constant (average) value and the fluctuating term, viz., $V_n(z)=\Av{V_n(z)}+\Delta V_n(z)$, the equations for functions $\psi_{\pm}(z|n)$ on the rough interval $\mathbb{L}$ may be written as
\begin{equation}\label{eq-psi-rough}
  \left[\frac{d^2}{d z^2}+\widetilde{k}_n^2-\Delta V_n(z) \right]\psi_{\pm}(z|n) = 0\ ,
\end{equation}
where $\widetilde{k}_n^2 = k_n^2-\Av{V_n(z)}$ is the newly defined mode energy which will be referred to as \emph{renormalized}. The renormalization is due to nonzero value of the quadratic in $\xi'(z)$ term in potential \eqref{Vn}, which on condition \eqref{Small_ampl} with fair accuracy equals
\begin{equation}\label{<V(n>}
  \Av{V_n(z)}\approx \left(1+\frac{\pi^2n^2}{3}\right)\left(\frac{\Xi}{w_0}\right)^2\ .
\end{equation}
Here, the dimensionless parameter $\Xi=\sqrt{\Av{\left[\xi'(z)\right]^2}}\sim \sigma/r_c$ is introduced to characterize the asperity sharpness. From the viewpoint of quantum dynamics, potential \eqref{<V(n>} may be regarded as the effective rectangular-shaped potential barrier which, in conjunction with modulating term $\Delta V_n(z)$, represents the particular implementation of the entropic barrier in the $n$-th channel of the conductor under investigation.

If parameter  $\Xi$ has a non-extremely small value, the mode energy, even though it is positive-valued in the smooth leads ($k_n^2>0$), in the rough section of the infinite wire can change the sign so that the inequality will hold true $\widetilde{k}_n^2<0$. This implies that a given mode, even if it would be extended in the conductor leads, within the rough section can be transformed into the evanescent mode, which does not carry noticeable current. Thus, in going from the smooth to the rough section of the conductor, the number of current-carrying modes reduces from $N_{c\,0}=[kw_0/\pi]$ (symbol $[\ldots]$ denotes the integer part of the enclosed number) to the value
\begin{equation}\label{N_c(Xi)}
  N_c(\Xi)=\left[\frac{1}{\pi}\sqrt{\frac{(kw_0)^2-\Xi^2}{1+\Xi^2/3}}\ \right]\ ,
\end{equation}
which is largely dependent on the degree of the asperity sharpness. For the mode of given index $n$ there exists a specific sharpness parameter value $\Xi= \Xi_n^{(\text{c-off})}$ the mode is ``cut off'' given that it is attained. This value is found from equation $\widetilde{k}_n^2=0$, being equal to
\begin{equation}\label{Cut-off_Xi}
  \Xi_n^{(\text{c-off})}=\frac{k_nw_0}{\sqrt{1+(\pi n)^2/3}}
  \qquad (k_n\geqslant 0)\ .
\end{equation}
At the critical value of parameter $\Xi$, which is equal to
\begin{equation}\label{Xi_cr}
  \Xi_{\text{cr}}=\Xi_1^{(\text{c-off})}=\sqrt{\frac{(kw_0/\pi)^2-1}{1/3+1/\pi^2}}\ ,
\end{equation}
the current transfer across the rough section becomes completely prohibited since the quantum waveguide turns into the cut-off state.

For modes that are extended in the rough conductor section we will seek functions $\psi_\pm(z|n)$ within the interval $\mathbb{L}$ as a sum of modulated exponent functions $\exp(\pm i \widetilde{k}_n z)$, whereas beyond this interval they will be sought as the superposition of non-renormalized exponentials $\exp(\pm i k_n z)$ balanced so as to obey BCs \eqref{BC_trial_GF},
\begin{equation}\label{psi_pm_pr}
\begin{aligned}
  & \psi_+^{(\mathrm{I})}(z|n) =\mathrm{e}^{i k_n (z+L/2)} + r_+^{(n)} \mathrm{e}^{-i k_n (z+L/2)}\ ,\\*
  & \psi_-^{(\mathrm{I})}(z|n) =t_-^{(n)} \mathrm{e}^{- i k_n (z+L/2)}\ ,\\
  & \psi_+^{(\mathrm{II})}(z|n) =\pi_+(z|n) \mathrm{e}^{i \widetilde{k}_n z} -
  i \gamma_+(z|n) \mathrm{e}^{-i \widetilde{k}_n z}\ ,\\*
  & \psi_-^{(\mathrm{II})}(z|n) =\pi_-(z|n) \mathrm{e}^{- i \widetilde{k}_n z} -
  i \gamma_-(z|n) \mathrm{e}^{i \widetilde{k}_n z}\ ,\\
  & \psi_+^{(\mathrm{III})}(z|n) = t_+^{(n)} \mathrm{e}^{i k_n (z-L/2)}\ ,\\*
  & \psi_-^{(\mathrm{III})}(z|n) = \mathrm{e}^{- i k_n (z-L/2)} + r_-^{(n)} \mathrm{e}^{i k_n (z-L/2)}\ .
\end{aligned}
\end{equation}
The relationship between the amplitude factors standing before exponentials in relationships~\eqref{psi_pm_pr} can be found from solution matching at points $z=\pm L/2$, considering that both functions $\psi_\pm(z|n)$ and their derivatives with respect to $z$ are continuous.

Factors $\pi_\pm(z|n)$ and $\gamma_\pm(z|n)$ entering into formulas~\eqref{psi_pm_pr} can be found with relative ease provided that the scattering induced by the fluctuating part of the potential can be regarded as a weak one. As the criterium for the scattering weakness we adopt the requirement for these factors to be smooth as against exponentials they multiply. This condition considerably facilitates consequential calculations, even if one considers the multiple scattering by the potential $\Delta V_n(z)$, since it enables us to go over from the second-order differential equations for functions $\pi_{\pm}(z|n)$ and $\gamma_{\pm}(z|n)$ to the first-order equations,
\begin{equation}\label{Pi_Gamma-dyn_eqs}
\begin{aligned}
  \pm\pi'_\pm(z|n) + i\eta(z|n) \pi_\pm(z|n) + \zeta_\pm^*(z|n)\gamma_\pm(z|n)&=0\ ,\\*
  \pm\gamma'_\pm(z|n) - i\eta(z|n)\gamma_\pm(z|n) + \zeta_\pm(z|n)\pi_\pm(z|n)&=0\ .
\end{aligned}
\end{equation}
Here, instead of the initial random function $\Delta V_n(z)$ we have introduced the effective spatially sub-averaged random fields
\begin{subequations}\label{Eta_Zeta-def}
\begin{align}
 \label{Eta-def}
  \eta(z|n) & =\frac{1}{2 \widetilde{k}_n}\int\limits_{z-l_n}^{z+l_n} \frac{d z'}{2 l_n} \Delta V_n(z')\ ,\\
 \label{Zeta-def}
  \zeta_\pm(z|n) & =\frac{1}{2 \widetilde{k}_n}\int\limits_{z-l_n}^{z+l_n} \frac{d z'}{2 l_n}
  \mathrm{e}^{\pm 2 i \widetilde{k}_n z'}\Delta V_n(z')\ .
\end{align}
\end{subequations}
The averaging interval $l_n$ in Eqs.~\eqref{Eta_Zeta-def} can be chosen arbitrary, yet obeying a pair of inequalities ${\widetilde{k}_n^{-1}\ll l_n\ll L_{sc}(n)}$, where $L_{sc}(n)$ is the characteristic spatial scale at which factors $\pi_{\pm}(z|n)$ and $\gamma_{\pm}(z|n)$ are varied. The BCs for equations \eqref{Pi_Gamma-dyn_eqs} result from matching the fields $\psi_\pm(z|n)$ and their derivatives at the end points $z=\pm L/2$,
\begin{subequations}\label{Pi_Gamma-init}
\begin{align}\label{Pi-init}
  \pi_\pm(\pm L/2|n) &=\frac{1}{2}\left(\frac{k_n}{\widetilde{k}_n}+1\right)t_\pm^{(n)}\mathrm{e}^{-i \widetilde{k}_n L/2}\ ,\\
 \label{Gamma-init}
  \gamma_\pm(\pm L/2|n) &=\frac{1}{2 i}\left(\frac{k_n}{\widetilde{k}_n}-1\right)t_\pm^{(n)}\mathrm{e}^{i \widetilde{k}_n L/2}\ .
\end{align}
\end{subequations}

The Wronskian of solutions \eqref{psi_pm_pr}, on condition of weak scattering, is calculated to
\begin{subequations}\label{Wronskian}
\begin{align}\label{W}
  &\mathcal{W}_n^{(\mathrm{I})}(z) = 2 i k_n t_-^{(n)}\ ,\\
  &\mathcal{W}_n^{(\mathrm{II})}(z) \approx 2 i \widetilde{k}_n \big[\pi_+(z|n)\pi_-(z|n) + \gamma_+(z|n)\gamma_-(z|n)\big]\ ,\\
  &\mathcal{W}_n^{(\mathrm{III})}(z) = 2 i k_n t_+^{(n)}\ .
\end{align}
\end{subequations}
In view of the first derivative absence in equation~\eqref{eq-psi-rough} this Wronskian is not $z$-dependent, wherefrom the equality $t_+^{(n)}=t_-^{(n)}=\mathfrak{T}^{(n)}$ follows, and with flow conservation condition \eqref{Conserv_mode} we also have the equality $|r_+^{(n)}|=|r_-^{(n)}|$. As is seen from definition \eqref{psi_pm_pr}, the quantity $\mathfrak{T}^{(n)}$ square modulus has the meaning of the mode $n$ transmission coefficient. But since this coefficient has appeared in the auxiliary problem, for the trial Green function instead of the exact one, we will further refer to it as the \emph{trial} transmission coefficient.

For subsequent calculations it is convenient to introduce functions
\begin{equation}\label{Gam}
  \Gamma_\pm(z|n) = \frac{\gamma_\pm(z|n)}{\pi_\pm(z|n)}\ ,
\end{equation}
which represent, according to \eqref{psi_pm_pr}, amplitude reflection coefficients for the modes excited at the ``plus''/``minus'' end of the conductor from corresponding intervals $(\pm L/2,z)$ ``filled'' with potential $\Delta V_n(z)$. These coefficients obey Riccati equations
\begin{equation}\label{Gam-equation}
  \pm\Gamma'_\pm(z|n)= 2i\eta(z|n)\Gamma_\pm(z|n)
  + \zeta_\pm^*(z|n) \Gamma_\pm^2(z|n)
  - \zeta_\pm(z|n)
\end{equation}
and boundary conditions
\begin{equation}\label{Gam-init}
 \Gamma_\pm(\pm L/2|n) =  - i \,\EuScript{R}_n \mathrm{e}^{i \widetilde{k}_n L}\ ,
\end{equation}
where $\EuScript{R}_n$ is the coefficient of quantum particle reflection from the potential step with different energy values, $k_n^2$ and $\widetilde{k}_n^2$, on its either sides,
\begin{equation}\label{Refl_pot_step}
  \EuScript{R}_n=\frac{k_n - \widetilde{k}_n}{k_n + \widetilde{k}_n}\ .
\end{equation}
Using functions $\pi_{\pm}(z|n)$, $\gamma_{\pm}(z|n)$ and $\Gamma_\pm(z|n)$, the trial Green function within the rough section of the conductor (i.\,e., for $z,z'\in\mathbb{L}$) in weak scattering can be given in the convenient matrix form
\begin{equation}\label{G-matrix}
  G^{(V)}_n(z,z') =\left(\mathrm{e}^{i \widetilde{k}_n z}; \mathrm{e}^{- i \widetilde{k}_n z}\right)
  \left(
  \begin{aligned}
  &G_{11}(z,z'|n)   &G_{12}(z,z'|n)\\
  &G_{21}(z,z'|n)   &G_{22}(z,z'|n)
  \end{aligned}
  \right)
  \left(
  \begin{aligned}
  &\mathrm{e}^{- i \widetilde{k}_n z'}\\
  &\mathrm{e}^{i \widetilde{k}_n z'}
  \end{aligned}
  \right)\ ,
\end{equation}
where matrix  $\big\|G_{ij}(z,z'|n)\big\|$ elements are constructed as
\begin{subequations}\label{G_ij(z,z')}
\begin{align}
\label{G_11(z,z')}
G_{11}(z,z'|n) & = \frac{\mathcal{A}_n(z')}{2 i \widetilde{k}_n}
\left[
\frac{\pi_+(z|n)}{\pi_+(z'|n)} \Theta(z-z') -
\frac{\gamma_-(z|n)}{\pi_-(z'|n)} \Gamma_+(z'|n) \Theta(z'-z)
\right]\ ,\\
\label{G_22(z,z')}
G_{22}(z,z'|n) & = \frac{\mathcal{A}_n(z')}{2 i \widetilde{k}_n}
\left[
\frac{\pi_-(z|n)}{\pi_-(z'|n)} \Theta(z'-z) -
\frac{\gamma_+(z|n)}{\pi_+(z'|n)} \Gamma_-(z'|n) \Theta(z-z')
\right]\ ,\\
\label{G_12(z,z')}
G_{12}(z,z'|n) & = - \frac{\mathcal{A}_n(z')}{2 \widetilde{k}_n}
\left[
\frac{\gamma_-(z|n)}{\pi_-(z'|n)} \Theta(z'-z) +
\frac{\pi_+(z|n)}{\pi_+(z'|n)} \Gamma_-(z'|n) \Theta(z-z')
\right]\ ,\\
\label{G_21(z,z')}
G_{21}(z,z'|n) & = - \frac{\mathcal{A}_n(z')}{2 \widetilde{k}_n}
\left[
\frac{\gamma_+(z|n)}{\pi_+(z'|n)} \Theta(z-z') +
\frac{\pi_-(z|n)}{\pi_-(z'|n)} \Gamma_+(z'|n) \Theta(z'-z)
\right]\ .
\end{align}
Factor $\mathcal{A}_n(z)$ in formulas~\eqref{G_ij(z,z')} is equal to
\begin{equation}\label{A_n(z)}
  \mathcal{A}_n(z)=\frac{1}{1+\Gamma_+(z|n)\Gamma_-(z|n)}\ .
\end{equation}
\end{subequations}

The Green function \eqref{G-matrix} and its elements \eqref{G_ij(z,z')} appear to be fully analogous to matrix Green function obtained previously in Ref.~\cite{Tar00}. However, in that paper the entropic barrier finite (on average) height was not taken into account because of the chosen model for the impurity potential. On examination of quantum transport in systems with rough boundaries the barrier is not modeled, but rather it is prescribed by the actual roughness structure. For this reason the impact of randomly (as well as regularly, see Ref.~\cite{GorTarShost13}) rough boundaries upon electron transport in wires of finite cross section is much more intricate and strong as compared with the influence of bulk irregularities.

In further analysis, along with function \eqref{G-matrix} which is defined within the conductor rough segment, the trial Green function will be required with the ``source'' point ($z'$) positioned at one end of the interval $\mathbb{L}$ whereas the observation point~($z$) on the other. In this case the trial propagator expression may be obtained directly from Eqs.~\eqref{Green-Cochi} and \eqref{psi_pm_pr}, and the result is as follows,
\begin{equation}\label{G-t}
  G^{(V)}_n(z,z') =\frac{\mathfrak{T}^{(n)}}{2 i k_n} \mathrm{e}^{i k_n (|z-z'|-L)}\ .
\end{equation}
From the constancy of Wronskian \eqref{Wronskian} we get the relationship
\begin{equation}\label{t_n}
  \mathfrak{T}^{(n)} =\frac{k_n}{\widetilde{k}_n}\cdot \frac{\mathcal{A}_n(z)}{\Pi_+(z|n)\Pi_-(z|n)}\ ,
\end{equation}
where $\Pi_\pm(z|n)$ are the functions defined through equality $\pi_\pm(z|n) = \mathfrak{T}^{(n)} \Pi_\pm(z|n)$. These functions obey the set of equations identical to \eqref{Pi_Gamma-dyn_eqs} and boundary conditions
\begin{equation}\label{Pi-initcond}
  \Pi_\pm(\pm L/2|n) =\frac{1}{2}\left(\frac{k_n}{\widetilde{k}_n}+1\right)\mathrm{e}^{-i \widetilde{k}_n L/2}\ .
\end{equation}

Expressions \eqref{G-matrix}--\eqref{G-t} enable one, in principle, to calculate any statistical moment of the trial Green function through the application of some quite non-trivial calculation technique whose details may be found, e.\,g., in Ref.~\cite{MakTar01}. In this study, where we solve the problem in space dimension larger than one, the evaluation of strictly 1D propagator $G^{(V)}_n(z,z')$ is, in a sense, an auxiliary problem. This enables us not to resort to the intricate calculation technique of \cite{MakTar01}, confining ourselves chiefly to qualitative calculations which we supplement, where necessary, by proper references.

\section{The impact of the intermode scattering}

In this section we examine the role of non-coherent scattering caused by intermode potentials \eqref{Unm}. The entire set of these potentials enter into equation \eqref{GDIAG-FIN} through mode-mixing operator $\hat{\mathsf{R}}$, whose matrix elements in coordinate-mode space  $\mathsf{M}$ are given by Eq.~\eqref{kernR}. Since this operator specifies completely the $T$-matrix structure, its properties substantially determine the possibility to simplify expression \eqref{T-oper} and thus to obtain the closed-form solution of master equation \eqref{GDIAG-FIN}.

\begin{figure}[h]
  \setcaptionmargin{.5in}%
  \centering
  \scalebox{.7}[.7]{\includegraphics{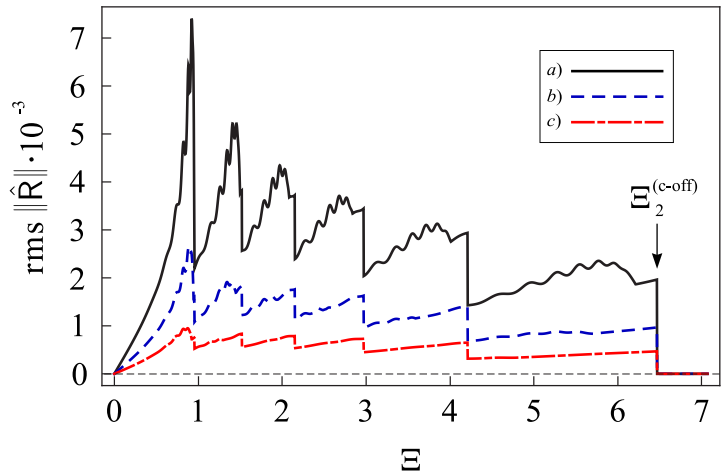}}
  \caption{The dependence of $\|\hat{\mathsf{R}}\|$ on the asperity sharpness, which is represented by formula Eq.~\eqref{R_norm-united} (the $\Xi$ value is varied at the cost of parameter $\sigma$ only). The conductor geometric parameters are chosen as follows: $L=2\, w_0$, $k=8\pi/w_0$, $L/r_c = 2000$ (\emph{a}), 1000 (\emph{b}), 500 (\emph{c}).
  \hfill\label{fig2}}
\end{figure}

The $T$-potential \eqref{T-oper} may be significantly simplified in two limiting cases, viz., for weak and strong mode intermixing. Taking account of this potential operator structure we will define the former case as the limit of small (as compared to unity) norm of the operator $\hat{\mathsf{R}}$, whereas the latter case will be specified as the limit of ${\|\hat{\mathsf{R}}\|\gg 1}$. The detailed estimation of operator $\hat{\mathsf{R}}$ norm is carried out in \ref{R_norm-estim}, and the result is depicted in Fig.~\ref{fig2}. As can be seen from the
graph, for the overwhelming majority of sharpness parameter values in subcritical range $0<\Xi<\Xi_{\text{cr}}$, where the quantum waveguide is in the conducting state, the roughness-induced intermode scattering is very strong. The scattering can be regarded as weak either if the asperities are extremely smooth (from estimate \eqref{R_norm-united}, this corresponds to enhanced inequality $\Xi\ll(1/\sqrt{N_{c0}})(r_c/L)\ll 1$) or in the range of fairly large $\Xi$'s, where one mode, at best, is allowed to propagate across the conductor rough section ($\Xi>\Xi_2^{(\text{c-off})}$).

In the parameter range where the operator $\hat{\mathsf{R}}$ norm exceeds substantially the unity value, the expression for potential \eqref{T-oper} may be transformed through the expansion of the inverse operator, $(\openone-\hat{\mathsf R})^{-1}$, in powers of the inverse mode-mixing operator $\hat{\mathsf{R}}^{-1}$,
\begin{align}\label{T_mu-R>1}
  {\mathcal T}_{n}=& \bm{P}_{n}\hat{\mathcal U}
  (\openone-\hat{\mathsf R})^{-1}\hat{\mathsf R}\bm{P}_{n}=
  \bm{P}_{n}\hat{\mathcal U}
  (\openone-\hat{\mathsf R})^{-1}\bm{P}_{n}
  = -\bm{P}_{n}\hat{\mathcal{U}}\hat{\mathsf{R}}^{-1}
  \big(\openone -\hat{\mathsf{R}}^{-1}\big)^{-1}\bm{P}_{n}
\notag\\
  =& -\sum_{k=0}^{\infty}\bm{P}_{n}\big[\hat{\mathcal{G}}^{(V)}\big]^{-1}
  \big(\hat{\mathsf{R}}^{-1}\big)^k\bm{P}_{n}
  =-\big[\hat{G}^{(V)}_{n}\big]^{-1}\sum_{k=0}^{\infty}\bm{P}_{n}
  \big(\hat{\mathsf{R}}^{-1}\big)^k\bm{P}_{n}\ .
\end{align}
Symbol  $\hat{\mathcal{G}}^{(V)}$ in Eq.~\eqref{T_mu-R>1} stands for the operator whose matrix in $\mathsf{M}$ is diagonally composed of trial Green function components,
\begin{equation}\label{G-matr}
  \bra{z,n}\hat{\mathcal{G}}^{(V)}\ket{z',m} =
  G_n^{(V)}(z,z')\delta_{nm}\ .
\end{equation}
Factor $\big[\hat{G}^{(V)}_{n}\big]^{-1}$ in Eq.~\eqref{T_mu-R>1} should also be interpreted in the operator sense, viz., as a continuous-variable operator which is the $n$-th diagonal element of the operator inverse to the operator $\hat{\mathcal{G}}^{(V)}$.

If we retain the term with $k=0$ only in the rhs of Eq.~\eqref{T_mu-R>1}, considering operator $\hat{\mathsf{R}}^{-1}$ norm to be small as compared to unity, we obtain the relationship
\begin{subequations}\label{Gnn->KG(V)n}
\begin{equation}\label{Gnn->G(V)n}
  G_{nn}(z,z')\approx \frac{1}{2}G_n^{(V)}(z,z')\ .
\end{equation}
In the specific case where only one conducting channel is left in consequence of sharpness-induced mode number reduction according to Eq.~\eqref{N_c(Xi)}, the only extended propagator that remains to be non-zero equals
\begin{equation}\label{Single-mode_case}
  G_{11}(z,z')\approx G_1^{(V)}(z,z')\ .
\end{equation}
\end{subequations}

The set of diagonal propagators $G_{nn}(z,z')$ specifies solely the ``diagonal'' part of the entire conductance. To calculate the whole conductance in the general case we also need non-diagonal propagators, which determine the transport between different ``entrance'' and ``exit'' channels. From Eq.~\eqref{Gmn->Gnn} it can be seen that on condition of strong mode intermixing, when inequality holds true $\|\hat{\mathsf{R}}\|\gg 1$, the estimation is valid
\begin{equation}\label{Gmn->(1/R)Gnn}
  |G_{mn}(z,z')|\sim\|\hat{\mathsf{R}}^{-1}\|\cdot|G_{nn}(z,z')|\ll |G_{nn}(z,z')|
  \qquad (m\neq n)\ ,
\end{equation}
wherefrom it can be deduced that in strong intermode scattering the conductance is mainly specified by the \emph{diagonal} transmission coefficients.

From Eqs.~\eqref{Gnn->KG(V)n} and \eqref{Gmn->(1/R)Gnn} the conclusion suggests itself that if the roughness-induced intermode scattering becomes strong in the sense of mode intermixing, there arises peculiar channel ``leveling''. The consequence of this is that to calculate the conductance in this case it suffices to have the trial propagators only. Mathematically, the leveling of the channels reduces to the appearance of factor $1/2$  on the right-hand-side of equality~\eqref{Gnn->G(V)n} instead of the factor close to unity, which is typical of weak mode intermixing.

\section{The conductance evaluation}

Given the system's Green function, the conductance may be evaluated either by means of linear response theory, through application of the well known Kubo formula \cite{Kubo57,KuboTodaHashitsume85}, or, alternatively, through the Landauer formalism \cite{Landauer57,ButtImryLandPinh85}. In paper Ref.~\cite{FisherLee81} it was proven that for infinitely long and homogeneous-on-average systems both of these approaches are essentially equivalent. Moreover, in Ref.~\cite{StoneSzafer88} the Landauer formula in its conventional form was directly obtained from the linear response theory. Meanwhile, it should be kept in mind that in the derivation of Kubo formula it is required that the current carriers in the finite system under consideration and in the attached leads be in the equilibrium. For this reason the application of Kubo theory is fully justified in the thermodynamic limit only~\cite{WuBerciu10}. In practical calculations of finite-system conductance, as pointed out in Ref.~\cite{Zirnbauer92}, the correct statement of boundary conditions is of paramount importance, which was long ago emphasized in the analysis of experiments with disordered mesoscopic conductors \cite{WashburnWebb86}.

From the foregoing considerations we choose to calculate the conductance through Landauer formula \eqref{Cond->t_mn}. According to \cite{StoneSzafer88}, the transmission coefficients are expressed in terms of the end-to-end Green propagators,
\begin{equation}\label{t_nm->G(L)}
  t_{mn}(L)=2i\sqrt{k_mk_n}G_{mn}\big(L/2,-L/2\big)\ .
\end{equation}
By neglecting the intermode Green function components when the mode intermixing is strong and considering the peculiar behavior of this function as the conductor is going over from multimode to single-mode regime (see Eqs.~\eqref{Gnn->KG(V)n}), we can reduce the conductance formula to the form
\begin{align}\label{g(L)->|t_n|2}
  g(L)= 
  g_0\sum_{n}\Big.^{\prime}k_n^2\big|G_n^{(V)}(L/2,-L/2)\big|^2=
  \frac{g_0}{4}\sum_{n}\Big.^{\prime}\big|\mathfrak{T}^{(n)}\big|^2\ .
\end{align}
Here, the prime at the sum symbol signifies that for $N_c(\Xi)=1$ the transmission coefficient must be doubled, which, in its turn, gives rise to the conductance quadrupling.

By expanding factor $\mathcal{A}_n(z)$ in Eq.~\eqref{t_n} in  Taylor series we can represent the transmission coefficient square modulus as the following double series,
\begin{align}\label{|t_n|2}
  |\mathfrak{T}^{(n)}|^2 =& \frac{k_n^2}{\widetilde{k}_n^2}
  \frac{1}{\big|\widetilde{\Pi}_+(z|n)\big|^2 \big|\widetilde{\Pi}_-(z|n)\big|^2}
  \sum_{l,m=0}^{\infty} (-1)^{l+m}
  \widetilde{\Gamma}_+^l(z|n) \widetilde{\Gamma}_+^{*m}(z|n)
  \widetilde{\Gamma}_-^l(z|n) \widetilde{\Gamma}_-^{*m}(z|n)
  \notag  \\*
  &\times
  \exp\left[ - 2 i (l-m) \int_{-L/2}^{L/2} \eta(z'|n) d z' \right]\ .
\end{align}
Here we have accomplished phase renormalization of functions ${\Pi}_\pm(z|n)$ and $\Gamma_\pm(z|n)$ according to the formulae
\begin{align}
\Pi_\pm(z|n) &=\widetilde{\Pi}_\pm(z|n) \exp
  \left[\pm i \int_z^{\pm L/2}   \eta(z_1|n) d z_1
  \right]\ ,\notag\\
\Gamma_\pm(z|n) &= \widetilde{\Gamma}_\pm(z|n) \exp\left[\mp 2 i \int_z^{\pm L/2}\eta(z_1|n) d z_1
  \right]\ .\notag
\end{align}
On averaging Eq.~\eqref{|t_n|2} we, firstly, should take into account the fact that exponential factor containing forward scattering field $\eta(z|n)$ can be averaged independently of other factors (see \ref{Mod_Trial_Green_func}). Besides, as the right-hand-side of Eq.~\eqref{|t_n|2} must not be $z$-dependent, we can assume the coordinate $z$ equal, say, to its value at the interval $\mathbb{L}$ right-hand side, using then the proper boundary conditions for functions $\widetilde{\Pi}_+(z|n)$ and $\widetilde{\Gamma}_+(z|n)$,
\begin{subequations}\label{Pi+Gamma+_BC+}
\begin{align}\label{Pi+BC+}
  \widetilde{\Pi}_+(L/2|n) & =\Pi_+(L/2|n)=
  \frac{1}{2}\left(\frac{k_n}{\widetilde{k}_n}+1\right)\mathrm{e}^{-i \widetilde{k}_n L/2}\\
  \label{Gamma+_BC+}
  \widetilde{\Gamma}_+(L/2|n) & =\Gamma_+(L/2|n)=- i \,\EuScript{R}_n \mathrm{e}^{i \widetilde{k}_n L}\ .
\end{align}
\end{subequations}
As a result, the transmittance \eqref{|t_n|2} upon averaging becomes
\begin{align}\label{|t_n|2(2)}
  \Av{|\mathfrak{T}^{(n)}|^2} = & \frac{4k_n^2}{\big|k_n+\widetilde{k}_n\big|^2}
  \sum_{l,m=0}^{\infty} i^{l-m}
  \EuScript{R}_n^l\EuScript{R}_n^{*m}
  \exp\left[i(l-m)\widetilde{k}_nL - (l-m)^2 \frac{2 L}{L_f^{(\text{eff})}(n)}\right]
  \notag\\*
  & \times \AV{\left|\widetilde{\Pi}_-(L/2|n)\right|^{-2}
  \widetilde{\Gamma}_-^l(L/2|n)\widetilde{\Gamma}_-^{*m}(L/2|n)}\ ,
\end{align}
where $L_f^{(\text{eff})}(n)$ is given by Eq.~\eqref{L_f,b-(eff)-def}. Allowing for the identity
\begin{equation}\label{Pi->Gamma}
  \frac{1}{\big|\widetilde{\Pi}_\pm(z|n)\big|^2} =
  \frac{\widetilde{k}_n}{k_n} \left[ 1 - \big|\widetilde{\Gamma}_\pm(z|n)\big|^2\right]\ ,
\end{equation}
which results from the flow conservation law \eqref{Const_det}, formula \eqref{|t_n|2(2)} for mode indices $n<N_c$ is transformed to
\begin{align}\label{|t_n|2(3)}
  \Av{|\mathfrak{T}^{(n)}|^2} = & \left(1 - \EuScript{R}_n^2\right)
  \sum_{l,m=0}^{\infty} i^{l-m}
  \EuScript{R}_n^l\EuScript{R}_n^{*m}
  \exp\left[i(l-m)\widetilde{k}_nL-2 (l-m)^2 \frac{L}{L_f^{(\text{eff})}(n)}\right]
  \notag\\*
  & \times \AV{\left[ 1 - \big|\widetilde{\Gamma}_-(L/2|n)\big|^2\right]
  \widetilde{\Gamma}_-^l(L/2|n)\widetilde{\Gamma}_-^{*m}(L/2|n)}\ .
\end{align}

Note that in the fictitious case where the fluctuating part of random potential $V_n(z)$ is formally disregarded the sums over indices $l$ and $m$ in Eq.~\eqref{|t_n|2(3)}, which originate from the series expansion of factor \eqref{A_n(z)} in both of the time-reversed propagators entering Eq.~\eqref{g(L)->|t_n|2}, would be completely uncoupled. By replacing the $\widetilde{\Gamma}_-$-function in Eq.~\eqref{|t_n|2(3)} with its ``initial'' value, $\widetilde{\Gamma}_-(L/2|n)=- i \,\EuScript{R}_n \exp\big(i\widetilde{k}_n L\big)$, we arrive at the formula
\begin{align}\label{|t_n|2(ball)}
  \Av{|\mathfrak{T}^{(n)}|^2}_\text{ball} =
  |\mathcal{Q}_n|^2\left(1 - \EuScript{R}_n^2\right)^2\ ,
\end{align}
which, from the standpoint of quantum mechanics, corresponds to the transmission through the rectangular barrier of height  $\Av{V_n(z)}$ and width $L$. Factor
\begin{equation}\label{Edge_res}
  \mathcal{Q}_n=\left[1-\left(\,\EuScript{R}_n \mathrm{e}^{i \widetilde{k}_n L}\right)^{2}\right]^{-1}
\end{equation}
in Eq.~\eqref{|t_n|2(ball)} describes the transmission resonances for quantum particle propagating through quasi-bounded states that arise in this conductor portion due to the mode mismatch at the ends of the corrugated section and the perfectly smooth leads.

\subsection{Disorder-induced localization}

To examine the role of the scattering related to random potential $\Delta V_n(z)$ entering equation~\eqref{eq-psi-rough} it is worthwhile to split the transmission coefficient \eqref{|t_n|2(3)} into conventionally ``diagonal'' and ``non-diagonal'' parts, the former containing the terms with $l=m$ whereas the latter incorporating the whole of the terms with $l\neq m$,
\begin{equation}\label{Transmittance(d+nd)}
  \Av{|\mathfrak{T}^{(n)}|^2}=\Av{|\mathfrak{T}^{(n)}|^2}_{\mathrm{d}}+\Av{|\mathfrak{T}^{(n)}|^2}_{\mathrm{nd}}\ .
\end{equation}
The ``diagonal'' transmittance has the form of the following one-fold series,
\begin{align}\label{|t_n|2(diag)}
  \Av{|\mathfrak{T}^{(n)}|^2}_{\mathrm{d}} = \left(1 - \EuScript{R}_n^2\right)
  \sum_{l=0}^{\infty}\EuScript{R}_n^{2l}
  \AV{\left|\widetilde{\Gamma}_-(L/2|n)\right|^{2l}\left[ 1 - \big|\widetilde{\Gamma}_-(L/2|n)\big|^2\right]}\ .
\end{align}
Its terms do not contain exponential factors, both oscillating and damping, which could reduce the transmittance value while being summed. From Eq.~\eqref{|t_n|2(diag)} it is seen that the averaging of this conductance portion is reduced to the calculation of correlator
\begin{equation}\label{Main_correlator}
  R_l(z|n)=\AV{\left|\widetilde{\Gamma}_-(z|n)\right|^{2l}}\qquad (l=0,1,2, \ldots)
\end{equation}
taken at $z=L/2$. The details of this correlator evaluation are given in {\ref{R_l(n)-calc}}, and as a result we arrive at the following expression for the first term on the right-hand side of Eq.~\eqref{Transmittance(d+nd)},
\begin{align}\label{|t_n|2(lambda)}
  \Av{|\mathfrak{T}^{(n)}|^2}_{\mathrm{d}} =& 2 \pi \int\limits_0^\infty  d\lambda\,  \frac{\lambda \tanh \pi \lambda}{\cosh \pi \lambda}
  \left[P_{-\frac{1}{2} + i \lambda}\left(\frac{1 + \EuScript{R}_n^2}{1 - \EuScript{R}_n^2}\right)\right]^2
  \exp\left[- \left(\frac{1}{4}+\lambda^2\right) \frac{L}{L_{b}^{(\text{eff})}(n)}\right]\ .
\end{align}
Here $P_{-\frac{1}{2}+i\lambda}(x)$ is the cone function \cite{AbrStegun64}.

From the result \eqref{|t_n|2(lambda)} it is not difficult to obtain the limiting values of the diagonal mode transmittance, which correspond to ``ballistic'' and ``localized'' regimes of quantum particle transport. In the former regime, that is consistent with inequality $L\ll L_{b}^{(\text{eff})}(n)$, the diagonal transmittance decreases linearly with the growth in the rough section length,
\begin{subequations}\label{T_n-asyptots}
\begin{equation}\label{T_n-ball}
  \Av{|\mathfrak{T}^{(n)}|^2}_\mathrm{d} \approx \frac{1 - \EuScript{R}_n^2}{1 + \EuScript{R}_n^2}
  \left[1-\frac{L}{L_{b}^{(\text{eff})}(n)}
  \frac{1 + \EuScript{R}_n^4}{(1 + \EuScript{R}_n^2)^2}\right]\ .
\end{equation}
In the opposite limit, where $L\gg L_{b}^{(\text{eff})}(n)$, the transmittance behavior changes to exponential fall, which is indicative of the one-dimensional Anderson localization characterized by the length of the flux diminution equal to quadruple backscattering length \cite{LifGredPast88},
\begin{align}\label{T_n-loc}
  \Av{|\mathfrak{T}^{(n)}|^2}_\text{d}
  \approx 2 \sqrt{\pi}  \left(1 - \EuScript{R}_n^2\right)
  K^2\left(\EuScript{R}_n\right) \left[\frac{L_{b}^{(\text{eff})}(n)}{L}\right]^{3/2}
  \exp\left[- \frac{L}{4 L_{b}^{(\text{eff})}(n)}\right]\ .
\end{align}
\end{subequations}
Factor $K(x)$ in this formula stands for the complete elliptic integral of the first kind \cite{AbrStegun64}.

\begin{figure}[h]
  \setcaptionmargin{.5in}%
  \centering
  \scalebox{.8}[.8]{\includegraphics{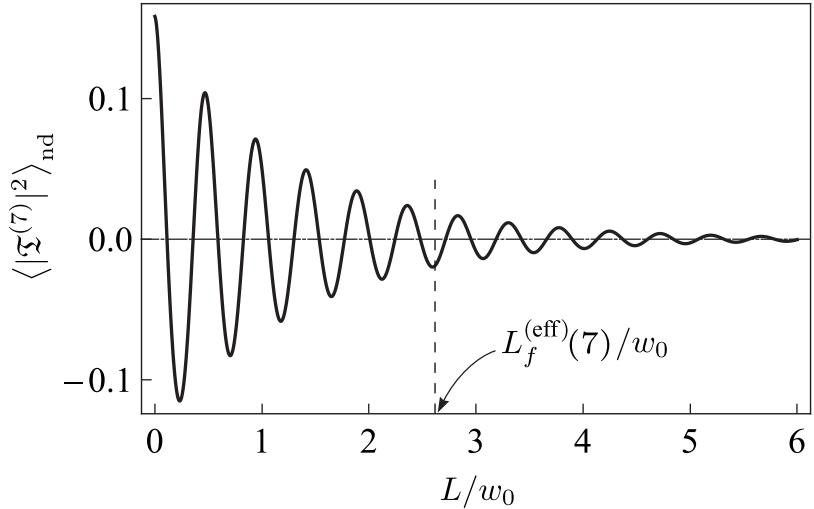}}
  \caption{The sketch of the ``non-diagonal'' transmittance dependence for the mode $n=7$ on the normalized length of the corrugated segment. The sharpness parameter $\Xi=0.8$, $k=8\pi/w_0$.
  \hfill\label{fig3}}
\end{figure}

\begin{figure}[h!]
  \setcaptionmargin{.5in}%
  \centering
  \scalebox{.8}[.8]{\includegraphics{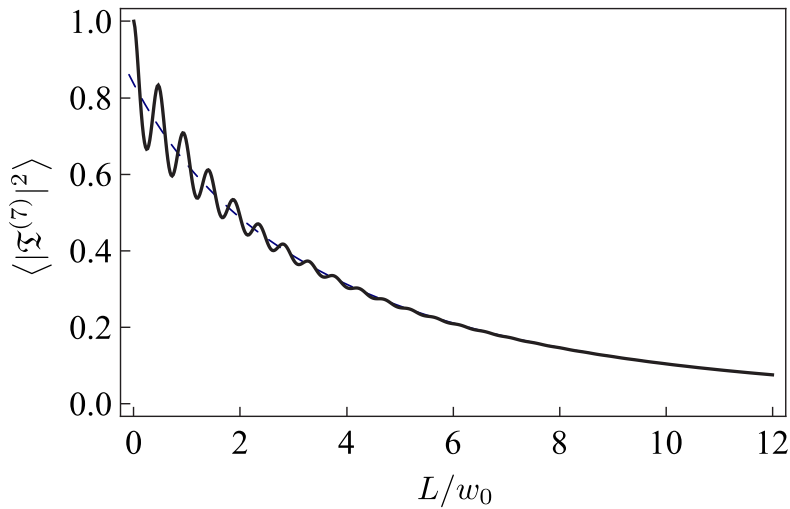}}
  \caption{Total transmittance dependence on the normalized length of the corrugated segment. The dashed line is the result given by Eq.~\eqref{|t_n|2(lambda)}. The parameters are the same as in Fig.~\ref{fig3}.
  \hfill\label{fig4}}
\end{figure}

As far as the ``non-diagonal'' part of the mode transmittance is concerned, it can be, in principle, calculated following the same procedure as the ``diagonal'' term given by Eq.~\eqref{|t_n|2(diag)}. Yet, this procedure, although allowing for multiple scattering by the potential $\Delta V_n(z)$, is quite involved and not necessary to be applied in this particular case. The point is that the exponential factor that couples the sums corresponding to time-reversed propagators in Eq.~\eqref{|t_n|2(3)}, even though it may attenuate rather slowly with the rough section length (when the forward scattering is so weak that the ratio $L/L_f^{(\text{eff})}(n)\ll 1$), yet disrupts the interference that results in Fabry--P\'erot oscillations described by the fictitious purely ballistic result~\eqref{|t_n|2(ball)}. In Fig.~\ref{fig3},
the result of numerical calculation of $\Av{|\mathfrak{T}^{(n)}|^2}_{\mathrm{nd}}$ is shown, that evidences its relatively small value and resonance behavior with the change in the length of the waveguide corrugated segment. Also, in Fig.~\ref{fig4} the summarizing graph of transmittance~\eqref{Transmittance(d+nd)} is presented, of which the first term, shown by dashed line, is calculated using integral expression~\eqref{|t_n|2(lambda)} whereas the second one is
obtained numerically straight from Eq.~\eqref{|t_n|2(3)}. It can be easily seen from this figure that most of the
transmittance value is provided by the subset of ``diagonal'' terms of the double series \eqref{|t_n|2(3)}. The oscillations of Fabry-P\'erot origin are relatively small and rapidly decreasing in amplitude in view of FP resonances suppression by the corrugation randomness.

Transition of the particular $n$-th mode conductance from ballistic form  \eqref{T_n-ball} to localized form \eqref{T_n-loc} with no conventional diffusion regime is a nontrivial fact for non-one-dimensional conducting systems. It has long been known that in strictly 1D surface-disordered systems two transport regimes only are allowed, viz., the ballistic and the localized regime~\cite{MakTar01}. The conventional diffusive regime characterized by the conductance decrease proportional to the inverse conductor length is impossible in such systems just because of their one-dimensional nature. This statement, however, is totally valid provided that the average value of the potential is strictly equal to zero. In this case there would be no mismatch between carrier states inside and outside the conductor rough section and, therefore, there should not take place FP-type resonances in the transmission coefficient.

As far as non-one-dimensional (multimode) system is concerned, this is not exactly the case. The availability of the term quadratic in $w'(z)$ in potential~\eqref{Vn}, which results essentially from inhomogeneity surface nature and is not on average equal to zero, makes it possible for the FP resonances to be revealed in the conductance. These resonances pertinent to the particular transverse mode are more pronounced in a~purely ballistic regime (with respect to the scattering due to potential $\Delta V_n(z)$), where formula~\eqref{|t_n|2(ball)} is applicable in full. As the disorder-related scattering is for all that taken into account, the resonances are gradually suppressed due to the mode incoherence that arises basically from the forward scattering and disappear completely when the given mode becomes Anderson-localized within the rough segment. In the former of two latter cases the mode transmittance is given by Eq.~\eqref{T_n-ball} while in the latter case it is presented by formula \eqref{T_n-loc}.

We thus have ascertained that for carriers of each distinct transverse-\-quan\-tiza\-tion mode of the multimode conductor or waveguide which includes finite segment with randomly rough lateral boundaries two transport regimes are normally feasible, viz., the quasi-ballistic regime, where the transmittance decreases on average linearly in the disordered section length, and the Anderson-localized regime, where the transmittance falls down exponentially. The prefix ``quasi-'' at the adjective ``ballistic'' is used to emphasize the fact that their exist particular deviations from the result obtained previously in Refs.~\cite{MakTar98,MakTar01} for one-mode (in a sense, strictly 1D) rough-bounded quantum wires. The deviations arise due to the mode mismatching at the rough section ends and reveal themselves in the form of Fabry-P\'erot-type oscillations of the transmittance.

The above pointed conduction regimes come into being non-simultaneously for the totality of extended modes in the multimode guiding system, which is consistent with the predictions of Refs.~\cite{SanchFreilYurkMarad98,SanchFreilYurkMarad99,RendMakIzr11} regarding the localization length hierarchy. As the regime conventionally termed the ``diffusive regime'' is concerned, where the transmittance is expected to scale inversely proportionally to the conductor length, it is challenging to substantiate this regime for multimode conductors and waveguides on the microscopic level without invoking some additional scattering (or, rather, dephasing) mechanisms.

\subsection{Entropic localization}

The analysis presented in the previous subsection relates mostly to the effect well-known in physics of random media, specifically, to the localization of Anderson type. The results we have obtained suggest that in randomly rough guiding systems, though non-one-dimensional in configurational space, this kind of localization reduces to localization peculiar to one-dimensional systems. This happens because the intermode scattering, which is essentially the non-1D effect, is rather strong for such systems in most of the roughness parameter region, as is demonstrated by Fig.~\ref{fig2}. Just due to this fact all the intermode propagators can be with asymptotic accuracy neglected in calculation of the entire transmittance (see the paragraph preceding Eq.~\eqref{g(L)->|t_n|2}), which may be regarded as the effective decomposition of the multimode rough-side conducting system into the set of almost independent single-mode quantum waveguides.

Yet, besides the conventional Anderson localization, one more type of localization is characteristic for rough-bounded guiding systems, which has purely entropic origin and is not related to the disorder \emph{per se}. The phenomenon is well-familiar in the classical waveguide theory and shows itself as the particular mode transformation from the mode of extended type to the evanescent mode, the latter being exponentially localized on the scale of its mode wavelength. In the system considered in this paper such a transformation occurs when the ``unperturbed'' mode energy $\widetilde{k}_n^2$ changes its sign, which happens when sharpness parameter $\Xi$ exceeds its particular-mode cut-off value, see~Eq.~\eqref{Cut-off_Xi}.

\begin{figure}[h!!]
  \setcaptionmargin{.5in}%
  \centering
  \scalebox{.8}[.8]{\includegraphics{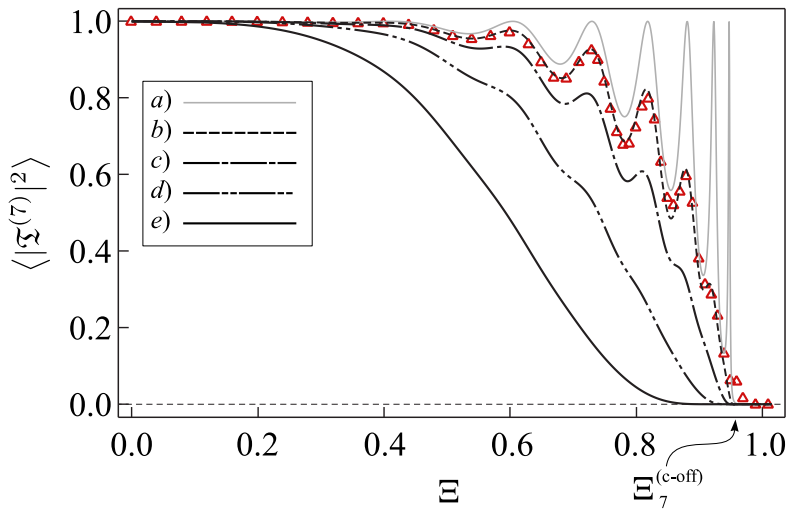}}
  \caption{The transmittance dependence on the sharpness parameter $\Xi$ for the rectangular barrier of height $\Av{V_n(z)}$ (curve~\emph{a}) and for modulated barrier $V_n(z)=\Av{V_n(z)} + \Delta V_n(z)$. The modulation is chosen to be consistent with geometric parameter $L/r_c = 3000$ (curve~\emph{b}), 1000 (curve~\emph{c}), 300 (curve~\emph{d}), and 80 (curve~\emph{e}); for all of the plots ${L = 2 w_0}$, $k=8\pi/w_0$. The triangles show the result of numerical simulations with the parameters corresponding to plot (\emph{b}) (the averaging is performed over 60 realizations of the roughness profile).
  \hfill\label{fig5}}
\end{figure}

To provide deeper insight into this kind of roughness-induced localization, in Fig.~\ref{fig5} we present the plots of %
the mean trial transmittance (coefficient $\av{|\mathfrak{T}^{(n)}|^2}$) for particular mode $n=7$, which are calculated numerically from Eq.~\eqref{|t_n|2(3)}.
The upper (gray-colored) curve demonstrates the dependence resulting from asymptotic expression~\eqref{|t_n|2(ball)}. Nominally, this formula is derived in the limit of $L/L_{f,b}^{(\text{eff})}(n)\to 0$, which implies the entire coherence of the given mode carriers as well as the lack of their Anderson localization within the rough conductor segment. In reality, though, such a dependence is somewhat idealized, since in its derivation we have neglected the fluctuating part of the potential entering Eq.~\eqref{G(V)_m-eq}, having kept its average value only. The transmittance shown by curve~(\emph{a}) corresponds virtually to the overbarrier propagation of a quantum particle with energy $k_7^2$ across the rectangular potential barrier of height $\Av{V_7(z)}$ and length $L$. The oscillations in this curve exhibit the resonance nature of such a transmission via quasi-bound states formed due to the interference of quantum waves repeatedly reflected from the barrier ends owing to the spectral mismatching of rough and smooth conductor sections (the FP effect).

Making allowance of the barrier modulation by function $\Delta V_n(z)$ leads to the damping of the FP resonances with the spatial rate equal to the sum of inverse ``slope'' and ``height'' forward scattering lengths, which are given by Eqs.~\eqref{Lf(s)(n)} and \eqref{Lf(h)(n)}. For sharpness parameter values $\Xi\gtrsim\sqrt{\sigma/w_0}$ the scattering lengths can be estimated as
\begin{align}\label{L_fb^(eff)-estim}
 & L_{f,b}^{(\text{eff})}(n)\sim\frac{\widetilde{k}_n^2}{r_c}
  \left(\frac{w_0}{\pi n}\right)^4 \frac{1}{\Xi^4}\ .
\end{align}
It is seen that as the parameter $\Xi$ approaches cut-off point $\Xi_n^{(\text{c-off})}$ from the left, where the mode is of extended nature, the localization length tends to decrease (formally to zero), so that in the vicinity of the cut-off point it becomes definitely smaller than the length of the rough conductor section. The transmittance resonance oscillations decrease in amplitude (curves (\emph{b}) to (\emph{e}) in Fig.~\ref{fig5}), going to zero with the growth in the correlation length at any fixed value of the sharpness parameter. In the limit of $r_c\to 0$, the transmittance proves to be of the same value and form as in the case of rectangular non-modulated potential barrier of height $\Av{V_n(z)}$ in spite of potential $V_n(z)$ possibly large fluctuations.

On reaching the cut-off point $\Xi=\Xi_7^{(\text{c-off})}$ the extended mode, whose transmittance is depicted in Fig.~\ref{fig5}, turns into the mode of evanescent type and thus becomes localized. The mechanism of localization in this case is fundamentally different from the Anderson mechanism in its conventional interpretation. By the Anderson localization is normally implied the formation of localized quantum or classical-wave states within the medium whose parameters are randomly varied in the space. Meanwhile, the waveguide mode cutting-off, i.\,e., their transformation from modes of extended type into the evanescent modes, is the effect of purely geometric nature, which is formally unrelated with any disorder. This sort of localization takes place not only in disordered waveguide systems similar to the one considered in this paper, but also in perfectly regular systems, in particular, in waveguides with periodically corrugated walls, which were studied in Ref.~\cite{GorTarShost13}.

\begin{figure}[h]
  \setcaptionmargin{.5in}%
  \centering
  \scalebox{.8}[.8]{\includegraphics{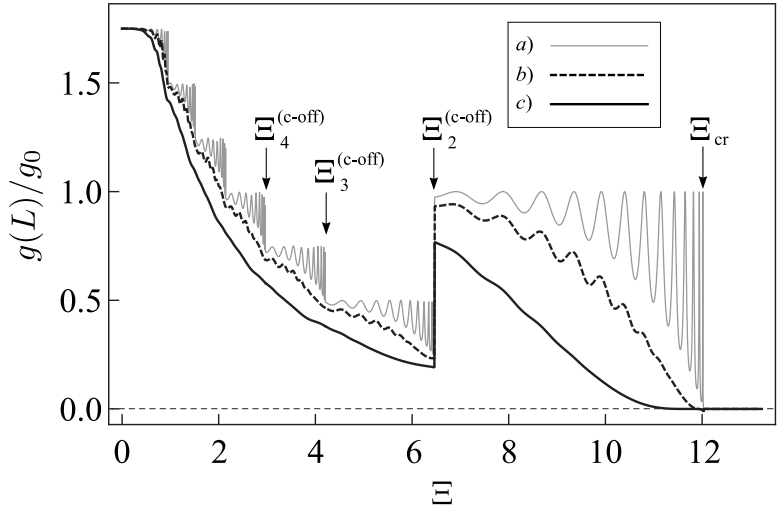}}
  \caption{The wire conductance vs the asperity sharpness degree. In the upper plot (\emph{a}), the gradient renormalization only (rectangular effective potential barriers $\Av{V_n(z)}$) is taken into account; the lower plots demonstrate the effect of the barrier random modulation by functions $\Delta V_n(z)$: ${L/r_c = 3000}$~(curve~\emph{b}), $L/r_c = 500$~(curve~\emph{c}); for all of the plots $L=2 w_0$, $k=8\pi/w_0$.
  \hfill\label{fig6}}
\end{figure}

In Fig.~\ref{fig6}, the dependence of full conductance~\eqref{g(L)->|t_n|2} on sharpness parameter $\Xi$ is shown.
The upper (gray) curve, which is depicted mainly for reference purposes, corresponds to the idealized case of $r_c\to 0$, where all of the extended modes are
subjected to solely average potentials and thus are definitely ballistic.\footnote{Of the same form is the asperity sharpness dependence of the wave conductance for waveguides with periodically corrugated walls \cite{GorTarShost13},  where all of the modes that do not fall into the Bloch spectrum forbidden regions are naturally ballistic.}
The envelope of this curve runs down stepwise, thereby reflecting the effect of sharpness-induced mode number reduction. While the conductance steps in ideal quantum contacts with smooth boundaries are equal exactly to one conductance quantum~$g_0$, for wires with randomly rough boundaries these steps prove to be exactly four times smaller in height, except for the last non-cut-off mode with $n=1$. This distinction results formally from the factor $1/2$ in the rhs of Eq.~\eqref{Gnn->G(V)n}, which originates from strong mode intermixing. In this situation the cutting of each of the extended modes causes the dimensionless conductance jump lower than the unity since on closing of each of the strongly entangled (and thus quasi-equivalent) waveguide modes the role of the conducting channels is homogeneously redistributed between all the remaining extended modes. At the same time, the contribution of the last non-cut-off mode ($n=1$, see Eq.~\eqref{Single-mode_case}) to the entire conductance is not thus reduced, as the carriers of this mode do not undergo the intermode scattering. Immediately prior to the conductor transition to the single-channel regime the conductance exhibits a peculiar dip, being in the single-channel regime considerably larger than two-channel and even three-channel conductance.

Such a conductance fall (more precisely, an increase in the resistivity) was already observed in the experiments with quantum wires (see, e.\,g., Refs.~\cite{Thornton89,Gilbertson11}) and recently with graphene nanoribbons \cite{Zozoulenko12}. This phenomenon, commonly referred to as the \emph{wire peak}, is routinely used to characterize the edge roughness of conductors under investigation. Yet, in spite of numerous observations of this effect its rigorous theory has not so far been available. In the theory we suggest in this paper the wire-peak effect manifests itself quite persistently, independently of the conductor being of regular or disordered nature. We believe that the persistency is explained by the universal nature of the phenomenon, specifically, by the transition of the considered quantum system to the single-mode regime, providing that in the pre-single-mode stage the intermode scattering (in fact, no matter where it originates from) is sufficiently strong.

\section{Conclusion}

To summarize, we have developed a unified theory for the conductance of an infinitely long multimode quantum wire whose finite segment has randomly rough lateral boundaries. The universality of the theory implies the possibility to take account of all feasible mechanisms of wave scattering in locally corrugated wires, both by contacts between the rough section and the infinite perfect leads and by the roughness of lateral boundaries within the same technical frameworks. We have managed to give an exact description of the latter type of scattering no matter how strong it is and what is the degree of the boundary asperity sharpness.

The rough finite-length insert into the initially smooth waveguiding system is shown to serve as a mode-specific randomly modulated potential barrier which can modify considerably the waveguide mode spectrum. The dominant role in the spectrum modification is played by the boundary asperity sharpness rather than by their amplitude. The sharpness degree governs not only the distortion of the particular mode lengthwise momentum but also the number of extended modes, i.\,e., the modes which are capable of traversing freely across the entire waveguide. Even though the roughness amplitude may be quite small and insufficient to influence noticeably the the extended mode quantity, as the asperities sharpen gradually, the number of conducting channels decreases monotonically, reaching zero when the governing sharpness parameter approaches some critical value. The reason for this is that with an increase in the sharpness of boundary asperities the roughness-induced potential barrier grows proportionally to the asperity mean square slope, which leads to the same effect as the waveguide geometric narrowing. On progressive narrowing of the waveguide, its extended modes are successively ``cut off'', in that they are transformed to ``evanescent'' waveguide modes which are localized exponentially at the scale of their mode wavelength. The conductance decrease that results from the roughness-induced reduction in the number of current-carrying modes can be regarded as a kind of localization,  which has geometric (or, equivalently, entropic) origin as opposed to Anderson (disorder-driven) localization.

In the course of the successive sharpness-induced mode cut-off the conductor, prior to being turned into the entirely ``locked out'' state, passes inevitably through the state where one extended mode only remains. This state is in a way unique since there are no other conducting channels for the carriers to be transferred to upon scattering. On the threshold of the one-mode stage the conductance, viewed as a function of the asperity sharpness, exhibits the local minimum. By comparing this unexpected conductance drop to the data known in the literature we suggest that it can be associated with the resistance \emph{wire peak} repeatedly observed experimentally in mesoscopic conductors, whose satisfactory theory has so far been unavailable.

Along with localization seen in the form of the sharpness-induced mode number reduction, the random nature of the waveguide corrugation results in additional kind of localization, which possesses all attributes of the Anderson localization  typical for one-dimensional quantum disordered systems. The 1D nature of this type of localization in guiding systems which, in reality, have non-one-dimensional structure can be explained as follows. The scattering induced by the surface roughness is, to the greatest extent, very strong provided that the asperities are not extremely smooth and/or the length of the waveguide rough section is not excessively small. Due to the great strength of the intermode scattering all the waveguide mode states appear to be strongly intermixed and thus effectively homogenized. This leads to the development of a new set of waveguide modes which can be regarded as being with asymptotic accuracy decoupled. For such 1D wave excitations it has long been established \cite{LifGredPast88,MakTar98,MakTar01} that under the action of random potentials they can propagate in ballistic or localized regimes only. The diffusive transport regime is prohibited for strictly 1D random systems. Thus, one of the important conclusions that stems from the results of the present paper is that in multimode waveguiding systems having inhomogeneity in the form of the rough-sided segment the waves can propagate either in the ballistic or in the localized regime, depending both on the particular mode transverse momentum and the length of the waveguide corrugated section.

In the domain of the sharpness-induced potential barrier, if one takes account of its average part only, the electron/wave transport proceeds in a resonant manner via quasi-steady states arising due to the interference of waveguide harmonics multiply scattered (owing to the propagation constant mismatch) between outer barrier edges. According to our theory this type of scattering should show up as Fabry-P\'erot-type oscillations of the conductance. These oscillations are more pronounced as the randomness-induced mode extinction length increases. With growing randomness of waveguide corrugation its propagating modes tend to become Anderson-localized within the rough segment, thereby inevitably violating the Fabry-P\'erot interference and suppressing the corresponding oscillations of the conductance.

Thus, in this study we have demonstrated that in corrugated conductors/waveguides with boundary asperities of quite small amplitude the carrier/wave localization can be provided by two independent, yet complementary, physical mechanisms, in contrast with conducting systems weakly disordered in the bulk. One type of localization, viz., localization of Anderson type, can be observed solely for waveguiding boundaries corrugated randomly. Yet another type of localization, which manifests itself through the cut-off effect well-known in the classical waveguide theory, is inherent both in randomly and in regularly corrugated quantum and classical waveguide-like systems. In our view, what is in fact intriguing and interesting is that the first of the above-mentioned kinds of localization is jointly controlled by the asperity sharpness degree and the roughness correlation dimension. At the same time the basic controlling parameter for the second, geometric type of localization is exclusively the degree of corrugation sharpness of the waveguide lateral boundaries.

\appendix

\section{The algebra of smooth ``amplitudes'' $\bm{\pi_\pm(z|n)}$ and $\bm{\gamma_\pm(z|n)}$ from Eqs.~(\ref{psi_pm_pr})}

Functions $\pi_\pm(z|n)$ and $\gamma_\pm(z|n)$ solve the \emph{evolutionary} problem given by equation set~\eqref{Pi_Gamma-dyn_eqs} and ``initial'' conditions~\eqref{Pi_Gamma-init}, which in fact are the conditions Eq.~\eqref{BC_trial_GF} ``pulled up'' from infinitely remote points to the corresponding ends of the rough waveguide section. Within the model~\eqref{psi_pm_pr} for the Cauchy problem sought for solution, the relation between constants  $r_{\pm}^{(n)}$ and $t_{\pm}^{(n)}$ is not primordially prescribed. To establish this relation, one can solve the evolutionary equations within segment $\mathbb{L}$, where the conductor side boundaries are subjected to deformation, and in this way to relate the values of $\pi_\pm(z|n)$ and $\gamma_\pm(z|n)$ at different ends of the rough conductor section.

The solution to Eqs.~\eqref{Pi_Gamma-dyn_eqs} can be obtained in the general form, without specifying the potentials. To this end, we perform simultaneous phase renormalization of functions $\pi_\pm(z|n)$ and $\gamma_\pm(z|n)$,
\begin{subequations}\label{pi-gamma-renorm}
\begin{align}\label{pi-renorm}
  \pi_\pm(z|n) &=\widetilde{\pi}_\pm(z|n) \exp
  \left[\pm i \int_z^{\pm L/2}   \eta(z_1|n) d z_1
  \right]
  \ ,\\
 \label{gamma-renorm}
  \gamma_\pm(z|n) &=\widetilde{\gamma}_\pm(z|n) \exp
  \left[\mp i \int_z^{\pm L/2}   \eta(z_1|n) d z_1
  \right]\ ,
\end{align}
\end{subequations}
and transform equations~\eqref{Pi_Gamma-dyn_eqs} to the form
\begin{subequations}\label{new-Pi_Gamma-dyn_eqs}
\begin{align}\label{new-Pi-dyn_eq}
  \pm\widetilde{\pi}'_\pm(z|n) + \widetilde{\zeta}_\pm^*(z|n)\widetilde{\gamma}_\pm(z|n)&=0\ ,\\*
 \label{new-Gamma-dyn_eq}
  \pm\widetilde{\gamma}'_\pm(z|n) + \widetilde{\zeta}_\pm(z|n)\widetilde{\pi}_\pm(z|n)&=0\ ,
\end{align}
\end{subequations}
where a pair of new random fields is introduced, viz.,
\begin{equation}\label{new-zeta}
  \widetilde{\zeta}_\pm(z|n) = \zeta_\pm(z|n) \exp
  \left[\pm 2 i \int_z^{\pm L/2}   \eta(z_1|n) d z_1
  \right]\ ,
\end{equation}
which possess the same correlation properties as the initial fields $\zeta_\pm(z|n)$.

Let us add to Eqs.~\eqref{new-Pi_Gamma-dyn_eqs} a set of complex conjugated equations and introduce two matrices of smooth amplitudes,
\begin{subequations}\label{Matrix_repr}
\begin{equation}\label{Ampl_matrix}
  \hat{\bm{I}}_+(z)=
  \begin{pmatrix}
  \widetilde{\pi}_+(z) & \widetilde{\gamma}_+(z)\\[3pt]
  \widetilde{\gamma}_+^*(z) & \widetilde{\pi}_+^*(z)
  \end{pmatrix}
  \ \text{,}\qquad
  \hat{\bm{I}}_-(z)=
  \begin{pmatrix}
  \widetilde{\pi}_-(z) & \widetilde{\gamma}_-^*(z)\\[3pt]
  \widetilde{\gamma}_-(z) & \widetilde{\pi}_-^*(z)
  \end{pmatrix}\ ,
\end{equation}
and two random-field matrices,
\begin{equation}\label{RandField_matrix}
  \hat{\bm{\zeta}}_+(z)=
  \begin{pmatrix}
  0 & \widetilde{\zeta}_+(z)\\[3pt]
  \widetilde{\zeta}_+^*(z) & 0
  \end{pmatrix}
  \ \text{,}\qquad
  \hat{\bm{\zeta}}_-(z)=
  \begin{pmatrix}
  0 & \widetilde{\zeta}_-^*(z)\\[3pt]
  \widetilde{\zeta}_-(z) & 0
  \end{pmatrix}\ .
\end{equation}
\end{subequations}
We temporarily omit index $n$ in definitions \eqref{Matrix_repr} for not to overload further expressions.

With notations~\eqref{Matrix_repr}, equations \eqref{new-Pi_Gamma-dyn_eqs} together with conjugated counterparts, may be cast in the matrix form,~viz.,
\begin{subequations}\label{Smooth_matr-eqs}
 \begin{align}\label{I+_eq}
  & \hat{\bm{I}}'_+(z)+\hat{\bm{I}}_+(z)\hat{\bm{\zeta}}_+(z)=0\ ,\\
  -& \hat{\bm{I}}'_-(z)+\hat{\bm{\zeta}}_-(z)\hat{\bm{I}}_-(z)=0\ .
 \end{align}
\end{subequations}
The solution to these equations can be represented in terms of $z$-ordered matrix exponentials,
\begin{equation}\label{I_pm(z)}
\begin{aligned}
  \hat{\bm{I}}_+(z) & =\hat{\bm{I}}_+(L/2)\,
  \hat{T}_z\exp\left[\int_z^{L/2}dz'\hat{\bm{\zeta}}_+(z')\right]\ ,\\
  \hat{\bm{I}}_-(z) & =
  \hat{T}_z\exp\left[\int_{-L/2}^z\!\!dz'\hat{\bm{\zeta}}_-(z')\right]
  \hat{\bm{I}}_-(-L/2)\ ,
\end{aligned}
\end{equation}
where $\hat{T}_z$ is the operator that arranges the multipliers in each of the terms of the exponential power
series in order of their coordinate arguments decreasing from left to right. From random-field matrices \eqref{RandField_matrix} being of zero-diagonal structure and from the well-known operator identity $\ln\det\hat{\mathbf{A}}=\Sp\ln\hat{\mathbf{A}}$ one can note that the determinants of the operator exponentials entering into Eqs.~\eqref{I_pm(z)} are all equal to unity, wherefrom the conclusion is made that those determinants are the $z$-independent functions,
\begin{equation}\label{Det_I=const}
  \det\hat{\bm{I}}_\pm(z)=\det\hat{\bm{I}}_\pm(\pm L/2)\ .
\end{equation}
By calculating these determinants using Eqs.~\eqref{Pi_Gamma-init} we arrive at the equality
\begin{equation}\label{Const_det}
  |\widetilde{\pi}_\pm(z|n)|^2-
  |\widetilde{\gamma}_\pm(z|n)|^2=|\mathfrak{T}^{(n)}|^2\frac{k_n}{\widetilde{k}_n}\ ,
\end{equation}
which is fulfilled identically for any extended mode, thereby enabling one to connect the values of function $\psi_{\pm}(z|n)$ at the corresponding entrance and exit points of the rough conductor section. To do this, one should take the values of functions $\pi_\pm(z|n)$ and $\gamma_\pm(z|n)$ at the section ends opposite to the sign index (viz., at point ${z=\mp L/2}$), which can be derived directly from Eq.~\eqref{psi_pm_pr}. This allows to express them in terms of trial reflection coefficients~$r_{\pm}^{(n)}$,
\begin{subequations}\label{Pi_Gamma-end-Rn}
\begin{align}
  \pi_\pm(\mp L/2|n) &=\frac{1}{2}\left(\frac{k_n}{\widetilde{k}_n}+1\right)
  \left(1 - r_\pm^{(n)}
  \EuScript{R}_n\right)
  \mathrm{e}^{i \widetilde{k}_n L/2}\ ,\\
  \gamma_\pm(\mp L/2|n) &=\frac{1}{2 i}
  \left(\frac{k_n}{\widetilde{k}_n}-1\right)
  \left(1 - r_\pm^{(n)}
  \EuScript{R}_n^{-1}\right)
  \mathrm{e}^{- i \widetilde{k}_n L/2}\ .
\end{align}
\end{subequations}
By substituting these values into the left-hand side of~\eqref{Const_det} we arrive at the equality
\begin{equation}\label{Conserv_mode}
  |\mathfrak{T}^{(n)}|^2+|r_{\pm}^{(n)}|^2=1\ ,
\end{equation}
which expresses the flux conservation law for one-dimensional system with finite-length inhomogeneity.

\section{The averaging of the trial Green function}
\label{Av_Trial_Green_func}

\subsection{The correlation properties of random fields~(\ref{Eta_Zeta-def})}

In the weak scattering limit, if the extra condition holds $r_c\ll L_{sc}(n)$, functions~\eqref{Eta_Zeta-def} may be regarded as approximately $\delta$-correlated random processes. This can be easily seen from direct evaluation of their correlators if one assumes the inequalities to meet
\begin{equation}\label{WS+short_corr}
  \widetilde{k}_n^{-1}, r_c\ll l_n\ll L_{sc}(n),L\ .
\end{equation}
Within these limitations the relationship between ``microscopic'' lengths, to which we relate the mode wavelength and the correlation radius, and ``macroscopic'' lengths $L_{sc}(n)$ and $L$ can be regarded as arbitrary. The calculation procedure for the correlators, though simple in essence, is rather tedious and cumbersome, so an interested reader can find its details in Ref.~\cite{MakTar98}. Here we will not dwell on the procedure \emph{per se}, but give the results of its application.

First, we represent potential $\Delta V_n(z)$ as a sum of ``slope'' and ``height'' terms, $\Delta V_n(z) = V_n^{(s)}(z) + V_n^{(h)}(z)$, which have, in view of Eq.~\eqref{Small_ampl}, the form
\begin{align}\label{Delta_Vn(s)}
  \Delta V_n^{(s)}(z) &=
  \left(1+\frac{\pi^2n^2}{3}\right) \frac{1}{w_0^2}
  \left[\xi'^2(z) - \Av{\xi'^2(z)}\right]\\
\label{Delta_Vn(h)}
  \Delta V_n^{(h)}(z) &=
  - 4 \frac{\pi^2 n^2}{w_0^3} \xi(z)\ .
\end{align}
The correlators of smoothed slope potentials $\eta_s(z|n)$ and $\zeta_{s\pm}(z|n)$ are calculated to
\begin{subequations}\label{Av_slope_fields}
\begin{align}
\label{<etaeta>s}
  &\Av{\eta_s(z|n)\eta_s(z'|n)} = \frac{1}{L_f^{(s)}(n)} F_{l_n}(z-z')\ ,\\
\label{<zetazeta+>s}
  &\Av{\zeta_{s\pm}(z|n)\zeta_{s\pm}^*(z'|n)} = \frac{1}{L_b^{(s)}(n)} F_{l_n}(z-z')\ ,\\
  &\Av{\zeta_{s\pm}(z|n)\eta_s(z'|n)} = \frac{1}{L_f^{(s)}(n)} f_{l_n}(z-z'|\widetilde{k}_n)
  \mathrm{e}^{\pm i \widetilde{k}_n (z+z')}\ ,\\
  &\Av{\zeta_{s\pm}(z|n)\zeta_{s\pm}(z'|n)} = \frac{1}{L_b^{(s)}(n)} f_{l_n}(z-z'|2 \widetilde{k}_n)
  \mathrm{e}^{\pm 2 i \widetilde{k}_n (z+z')}\ ,
\end{align}
\end{subequations}
where
\begin{subequations}\label{L_f-b}
\begin{align}
\label{Lf(s)(n)}
  \frac{1}{L_f^{(s)}(n)} &= \frac{1}{2 \widetilde{k}_n^2} \left(1+\frac{\pi^2n^2}{3}\right)^2
  \left(\frac{\sigma}{w_0}\right)^4
  \int\limits_{-\infty}^\infty \frac{d q}{2 \pi}  q^4 \widetilde{W}^2(q)\ ,\\
\frac{1}{L_b^{(s)}(n)} &= \frac{1}{2 \widetilde{k}_n^2} \left(1+\frac{\pi^2n^2}{3}\right)^2
  \left(\frac{\sigma}{w_0}\right)^4
  \int\limits_{-\infty}^\infty \frac{d q}{2 \pi} \big(q^2 -  \widetilde{k}_n^2\big)^2
  \widetilde{W}(q + \widetilde{k}_n) \widetilde{W}(q - \widetilde{k}_n)
\end{align}
\end{subequations}
are the inverse lengths of forward ($f$) and backward ($b$) scattering related to these potentials, $\widetilde{W}(q)$ is the Fourier transform of function $W(z)$ from Eq.~\eqref{Xi_bin_corr}. The functions
\begin{subequations}\label{F_ln_f_ln}
\begin{align}
 & F_{l_n}(z) =  \frac{1}{2 l_n} \left(1-\frac{|z|}{2 l_n}\right) \Theta (1-|z|/2 l_n)\ ,
\label{F_ln}\\
 & f_{l_n}(z|\widetilde{k}_n) =
  \frac{1}{4 \widetilde{k}_n l_n^2}
  \sin\left[2 \widetilde{k}_n l\left(1-\frac{|z|}{2 l_n}\right)\right] \Theta (1-|z|/2 l_n)
\label{f_ln}
\end{align}
\end{subequations}
possess the property that if they are integrated over any interval of length $|\Delta z|\gg l_n$, the function \eqref{F_ln} behaves like a true $\delta$-function whereas the result of function \eqref{f_ln} integration prove to be parametrically small ($\sim 1/\widetilde{k}_n l_n\ll 1$). For this reason, among all the correlators \eqref{Av_slope_fields} correlators \eqref{<etaeta>s} and \eqref{<zetazeta+>s} only may be regarded as having non-zero value, whereas the other two can reasonably be omitted.

The analogous calculation of ``height'' field correlators, which are related to potential~\eqref{Delta_Vn(h)}, is expressed as
\begin{subequations}\label{Av_height_fields}
\begin{align}
\label{<etaeta>h}
  &\Av{\eta_h(z|n)\eta_h(z'|n)} =
  \frac{1}{L_f^{(h)}(n)} F_{l_n}(z-z')\ ,\\
\label{<zetazeta+>h}
  &\Av{\zeta_{h\pm}(z|n)\zeta_{h\pm}^*(z'|n)} =
  \frac{1}{L_b^{(h)}(n)} F_{l_n}(z-z')\ ,
\end{align}
\end{subequations}
where the inverse forward and backward scattering lengths are given by
\begin{subequations}\label{L_f-b(h)}
\begin{align}
\label{Lf(h)(n)}
  \frac{1}{L_f^{(h)}(n)} &= 4 \frac{\pi^4 n^4 \sigma^2}{w_0^6 \widetilde{k}_n^2} \widetilde{W}(0)\ ,\\
\label{Lb(h)(n)}
  \frac{1}{L_b^{(h)}(n)} &= 4 \frac{\pi^4 n^4 \sigma^2}{w_0^6 \widetilde{k}_n^2}
  \widetilde{W}(2 \widetilde{k}_n)\ .
\end{align}
\end{subequations}
%

\subsection{The averaging of function~(\ref{G-matrix})}
\label{Mod_Trial_Green_func}

The elements of the smooth amplitude matrix in Eq.~\eqref{G-matrix} are constructed of causal-type functionals, hence, for their averaging the Furutsu-Novikov method can be applied \cite{Klyatskin05}. As an important component of this method, dynamic equations~\eqref{Pi_Gamma-dyn_eqs} and \eqref{Gam-equation} are of particular use, which specify the functional dependence of primary ``building blocks'' $\pi_{\pm}(z|n)$, $\gamma_{\pm}(z|n)$ and $\Gamma_\pm(z|n)$ on the effective random fields~\eqref{Eta_Zeta-def}. Since in weak scattering the correlation between random field $\eta(z|n)$ and $\zeta_\pm(z|n)$ is negligibly small, it is convenient to phase-renormalize functions $\pi_{\pm}(z|n)$ and $\gamma_{\pm}(z|n)$ according to Eqs.~\eqref{pi-gamma-renorm}, thus obtaining more simple dynamic equations for renormalized functions $\widetilde{\pi}_{\pm}(z|n)$ and $\widetilde{\gamma}_{\pm}(z|n)$. Functions $\Gamma_\pm(z|n)$ are renormalized respectively,
\begin{equation}\label{Gam_renorm}
  \Gamma_\pm(z|n) = \widetilde{\Gamma}_\pm(z|n) \exp\left[\mp 2 i \int_z^{\pm L/2}\eta(z_1|n) d z_1
  \right]\ ,
\end{equation}
so for new blocks $\widetilde{\Gamma}_\pm(z|n)$ the equations hold true
\begin{equation}\label{Gam_eq}
  \pm\widetilde{\Gamma}'_\pm(z|n)=- \widetilde{\zeta}_\pm(z|n)
  + \widetilde{\zeta}_\pm^*(z|n) \widetilde{\Gamma}_\pm^2(z|n)\ .
\end{equation}
Boundary conditions for blocks marked by tilde sign coincide with BCs for initial, non-renormalized blocks, whereas formulas~\eqref{G_ij(z,z')} in terms of renormalized functionals are written~as
\begin{subequations}\label{tild_G_ij(z,z')}
\begin{align}
  G_{11}(z,z'|n) = & \frac{\mathcal{A}_n(z')}{2 i \widetilde{k}_n}
  \exp\bigg(-i\int\limits_{z'}^z\eta(z_1|n)dz_1\bigg)
  \Bigg[\Theta(z-z')\frac{\widetilde{\pi}_+(z|n)}{\widetilde{\pi}_+(z'|n)}\notag \\*
  & - \Theta(z'-z)\frac{\widetilde{\gamma}_-(z|n)}{\widetilde{\pi}_-(z'|n)} \widetilde{\Gamma}_+(z'|n)
  \exp\bigg(-2i\int\limits_{-L/2}^{L/2}\eta(z_1|n)dz_1\bigg)
  \Bigg]\ ,
  \label{tild_G_11(z,z')}
\\
  G_{22}(z,z'|n) = & \frac{\mathcal{A}_n(z')}{2 i \widetilde{k}_n}
  \exp\bigg(i\int\limits_{z'}^z\eta(z_1|n)dz_1\bigg)
  \Bigg[ \Theta(z'-z)\frac{\widetilde{\pi}_-(z|n)}{\widetilde{\pi}_-(z'|n)}  \notag\\*
  & - \Theta(z-z') \frac{\widetilde{\gamma}_+(z|n)}{\widetilde{\pi}_+(z'|n)} \widetilde{\Gamma}_-(z'|n)
  \exp\bigg(-2i\int\limits_{-L/2}^{L/2}\eta(z_1|n)dz_1\bigg)
  \Bigg]\ ,
  \label{tild_G_22(z,z')}
\\
  G_{12}(z,z'|n) = & - \frac{\mathcal{A}_n(z')}{2 \widetilde{k}_n}
  \exp\bigg(-i\int\limits_{-L/2}^{z}\eta(z_1|n)dz_1-  i\int\limits_{-L/2}^{z'}\eta(z_1|n)dz_1\bigg)\notag\\*
  & \times
  \Bigg[ \Theta(z'-z)\frac{\widetilde{\gamma}_-(z|n)}{\widetilde{\pi}_-(z'|n)}  +
  \Theta(z-z')\frac{\widetilde{\pi}_+(z|n)}{\widetilde{\pi}_+(z'|n)} \widetilde{\Gamma}_-(z'|n)
  \Bigg]\ ,
  \label{tild_G_12(z,z')}
\\
  G_{21}(z,z'|n)= & - \frac{\mathcal{A}_n(z')}{2 \widetilde{k}_n}
  \exp\bigg(-i\int\limits_{z}^{L/2}\eta(z_1|n)dz_1-  i\int\limits_{z'}^{L/2}\eta(z_1|n)dz_1\bigg)\notag\\*
  & \times
  \Bigg[\Theta(z-z')\frac{\widetilde{\gamma}_+(z|n)}{\widetilde{\pi}_+(z'|n)}  +
  \Theta(z'-z)\frac{\widetilde{\pi}_-(z|n)}{\widetilde{\pi}_-(z'|n)} \widetilde{\Gamma}_+(z'|n)
  \Bigg]\ ,
  \label{tild_G_21(z,z')}
\end{align}
\begin{equation}\label{An(z)tilde}
  \mathcal{A}_n(z)=\Bigg[1+\widetilde{\Gamma}_+(z|n)\widetilde{\Gamma}_-(z|n)
  \exp\bigg(-2i\int\limits_{-L/2}^{L/2}\eta(z_1|n)dz_1\bigg)\Bigg]^{-1}\ .
\end{equation}
\end{subequations}

Now we detail the averaging procedure for the first term in Eq.~\eqref{tild_G_11(z,z')} as an example. The remaining elements of matrix~\eqref{tild_G_ij(z,z')} are averaged in a similar way. The matrix elements~\eqref{tild_G_ij(z,z')} are dependent upon two spatial variables. However, for the application of Furutsu-Novikov method the equations for functionals subject to averaging have to be ordinary rather than partial differential equations. This can be achieved if matrix elements~\eqref{tild_G_ij(z,z')} are firstly Fourier-transformed over one of the variables and then inversely transformed at the last stage.

The Fourier transform over $z$ of the first term in Eq.~\eqref{tild_G_11(z,z')}, after expansion of factor $\mathcal{A}_n(z)$ into the power series, takes on the form
\begin{align}\label{G_11(+)(q,z)}
  \widetilde{G}_{11}^{(1)}(q,z'|n)= & \frac{\mathrm{e}^{-iqz'}}{2i\widetilde{k}_n}
  \sum_{l=0}^{\infty}(-1)^l \exp\Bigg[-2il\int\limits_{-L/2}^{L/2}\eta(z_1|n)dz_1
  \Bigg]
  \notag\\*
  & \times
  \int\limits_{z'}^{L/2}dz\frac{\widetilde{\pi}_+(z|n)}{\widetilde{\pi}_+(z'|n)}
  \widetilde{\Gamma}_+^l(z'|n) \widetilde{\Gamma}_-^l(z'|n)
  \exp\Bigg[-iq(z-z')-i\int\limits_{z'}^{z}\eta(z_1|n)dz_1\Bigg]
  \ .
\end{align}
Since random fields~\eqref{Eta_Zeta-def} are approximately Gaussian distributed, the exponentials containing the field $\eta(z|n)$ are averaged exactly \cite{Vasilyev76},
\begin{align}\label{Averaged_forv_exp}
  \bigg<\exp\bigg(-2il\int_{-L/2}^{L/2}\eta(z_1|n)dz_1
      -i&\int_{z'}^{z}\eta(z_1|n)dz_1\bigg)\bigg>
  \notag\\*
  &=
  \exp\left\{-\frac{2}{L_f^{(\text{eff})}(n)}
  \Big[Ll^2+\big(l+1/4\big)(z-z')\Big]\right\}\ .
\end{align}
Here we use the notation
\begin{equation}\label{L_f,b-(eff)-def}
  \frac{1}{L_{f,b}^{(\text{eff})}(n)}=
  \frac{1}{L_{f,b}^{(s)}(n)}+\frac{1}{L_{f,b}^{(h)}(n)}\ .
\end{equation}

The remaining factors in Eq.~\eqref{G_11(+)(q,z)} are the functionals of random fields  $\widetilde{\zeta}_\pm(z|n)$, the functionals labeled by index ``$+$'' being dependent on field $\widetilde{\zeta}_+(z|n)$ and its conjugated counterpart, whereas the ``minus''-labeled functionals contain the random fields labeled by ``$-$'' sign. The fields labeled by different sign indices, in view of their being effectively $\delta$-correlated, are not subjected to pairing, so the average product of functionals of different ``signs'' is decoupled into the product of averages.

Consider the average value of the last factor in the rhs of Eq.~\eqref{G_11(+)(q,z)},
\begin{equation}\label{Gamma(-)^l-av}
  \EuScript{T}_-^{(l)}(z|n)=\AV{\widetilde{\Gamma}_-^l(z|n)}\ .
\end{equation}
The function subject to averaging here obeys the equation
\begin{subequations}\label{Gamma(l)-eq+init}
\begin{equation}\label{Gamma(l)-eq}
  \frac{d\widetilde{\Gamma}_-^l(z|n)}{dz}=l\Big[\widetilde{\zeta}_-(z|n)\widetilde{\Gamma}_-^{l-1}(z|n)-
  \widetilde{\zeta}_-^*(z|n)\widetilde{\Gamma}_-^{l+1}(z|n)\Big]
\end{equation}
and the ``initial'' condition
\begin{equation}\label{Gamma(l)-init}
  \widetilde{\Gamma}_-^l(-L/2|n)=\left(- i \,\EuScript{R}_n \mathrm{e}^{i \widetilde{k}_n L}\right)^l\ .
\end{equation}
\end{subequations}
The correlators that arise in the rhs of Eq.~\eqref{Gamma(l)-eq} in the course of averaging are reduced, with the aid of Furutsu--Novikov formalism, to the following form,
\begin{subequations}\label{<zetaGamma><zeta*Gamma>}
 \begin{align}
  \label{<zetaGamma>}
   \left<\widetilde{\zeta}_-(z|n)\widetilde{\Gamma}_-^{l-1}(z|n)\right>= &
   \frac{1}{2 L_{b}^{(\text{eff})}(n)}
   \Bigg<\frac{\delta\widetilde{\Gamma}_-^{l-1}(z|n)}{\delta\widetilde{\zeta}_-^*(\varsigma|n)}\Bigg>
   \Bigg|_{\varsigma=z-0}\ ,\\[6pt]
  \label{<zeta*Gamma>}
   \left<\widetilde{\zeta}_-^*(z|n)\widetilde{\Gamma}_-^{l+1}(z|n)\right>= &
   \frac{1}{2 L_{b}^{(\text{eff})}(n)}
   \Bigg<\frac{\delta\widetilde{\Gamma}_-^{l+1}(z|n)}{\delta\widetilde{\zeta}_-(\varsigma|n)}\Bigg>
   \Bigg|_{\varsigma=z-0}\ .
 \end{align}
\end{subequations}
The averages on the right-hand-sides of Eqs.~\eqref{<zetaGamma><zeta*Gamma>} are evaluated by applying the technique set forth in Ref.~\cite{KanerTar88},
\begin{subequations}\label{VarDerivatives}
 \begin{align}
  \label{VarDeriv_1}
   \Bigg<\frac{\delta\widetilde{\Gamma}_-^{l-1}(z|n)}{\delta\widetilde{\zeta}_-^*(\varsigma|n)}\Bigg>
   \Bigg|_{\varsigma=z-0}=& -(l-1)\left<\widetilde{\Gamma}_-^{l}(z|n)\right>\ ,\\
  \label{VarDeriv_2}
   \Bigg<\frac{\delta\widetilde{\Gamma}_-^{l+1}(z|n)}{\delta\widetilde{\zeta}_-(\varsigma|n)}\Bigg>
   \Bigg|_{\varsigma=z-0}=& (l+1)\left<\widetilde{\Gamma}_-^{l}(z|n)\right>\ ,
 \end{align}
\end{subequations}
whereupon the equation for function~\eqref{Gamma(-)^l-av} is readily obtained
\begin{equation}\label{Gamma(-)^l-av-EQ}
  \frac{d\EuScript{T}_-^{(l)}(z|n)}{dz}=
  - \frac{l^2}{L_{b}^{(\text{eff})}(n)}\EuScript{T}_-^{(l)}(z|n)\ .
\end{equation}
Its solution yields
\begin{equation}\label{Gamma(-)^l-av-SOL}
  \AV{\widetilde{\Gamma}_-^l(z|n)}=
  \left(- i \,\EuScript{R}_n \mathrm{e}^{i \widetilde{k}_n L}\right)^l
  \exp\left[-\frac{l^2}{L_{b}^{(\text{eff})}(n)}\big(L/2+z\big)\right]\ .
\end{equation}
The expression for function $\widetilde{\Gamma}_+(z|n)$ moments is obtained in a similar way, being different from Eq.~\eqref{Gamma(-)^l-av-SOL} by the sign at the variable $z$ in the exponential.

The last functional to be averaged in formula~\eqref{G_11(+)(q,z)} is
\begin{equation}\label{Q11+l(z|n)}
  Q_{11+}^{(l)}(z'|n)=  \widetilde{\Gamma}_+^l(z'|n)
  \int\limits_{z'}^{L/2}dz\frac{\widetilde{\pi}_+(z|n)}{\widetilde{\pi}_+(z'|n)}
  \exp\left\{- \left[iq+\frac{2 l+1/2}{ L_{f}^{(\text{eff})}(n)}\right]
  (z-z')\right\}\ .
\end{equation}
The equation it satisfies is as follows,
\begin{align}\label{Q11+l(z|n) - equation}
  \frac{d}{d z} Q_{11+}^{(l)}(z|n) =& \Big[
  (l+1)\widetilde{\zeta}_+^*(z|n) \widetilde{\Gamma}_+(z|n)
  - l \widetilde{\zeta}_+(z|n) \widetilde{\Gamma}_+^{-1}(z|n) + \mu_{n+}^{(l)}(q)\Big] Q_{11+}^{(l)}(z|n)
  - \widetilde{\Gamma}_+^l(z|n)\ ,
\end{align}
where the notation is used
\begin{equation}
  \mu_{n+}^{(l)}(q) = iq+\frac{1}{ L_{f}^{(\text{eff})}(n)}(2l+1/2)\ .
\end{equation}
Upon averaging Eq.~\eqref{Q11+l(z|n) - equation} we arrive at the equation
\begin{equation}\label{Q11+l(z|n)-av - equation}
  \frac{d}{d z}\AV {Q_{11+}^{(l)}(z|n)} =
  \left[ \mu_{n+}^{(l)}(q) + \frac{(l+1)^2 + l^2}{2  L_{b}^{(\text{eff})}(n)}  \right]
  \AV{Q_{11+}^{(l)}(z|n)} -
  \AV{\widetilde{\Gamma}_+^l(z|n)}\ ,
\end{equation}
whose solution that satisfies the obvious zero condition at the right end of the interval $\mathbb{L}$ is
\begin{align}\label{AV_Q_11+}
  \AV {Q_{11+}^{(l)}(z'|n)}= & \left(- i \,\EuScript{R}_n \mathrm{e}^{i \widetilde{k}_n L}\right)^l
  \frac{\exp\left[-l^2\big(L/2-z'\big)\big/ L_{b}^{(\text{eff})}(n)\right]}
  {iq+\big(l+1/2\big)\big/ L_{b}^{(\text{eff})}(n)+
  \big(2l+1/2\big)\big/ L_{f}^{(\text{eff})}(n)} \notag\\*[3pt]
  &\times\left\{1-\exp\left[-\bigg(iq+
  \frac{2l+1/2}{L_{f}^{(\text{eff})}(n)}
  +\frac{l+1/2}{L_{b}^{(\text{eff})}(n)}\bigg)
  \left(L/2-z'\right)\right] \right\}\ .
\end{align}
With results \eqref{Averaged_forv_exp}, \eqref{Gamma(-)^l-av-SOL} and \eqref{AV_Q_11+}, the average value of the quantity in rhs of~\eqref{G_11(+)(q,z)} is expressed as
\begin{align}\label{G_11(q)}
  \AV{\widetilde{G}_{11}^{(1)}(q,z'|n)}
  =& \frac{\mathrm{e}^{-iqz'}}{2i\widetilde{k}_n}  \sum_{l=0}^{\infty}
  \left(\,\EuScript{R}_n \mathrm{e}^{i \widetilde{k}_n L}\right)^{2 l}
  \left[ iq+\frac{2l+1/2}{L_{f}^{(\text{eff})}(n)}
  +\frac{l+1/2}{L_{b}^{(\text{eff})}(n)}\right]^{-1}
\notag\\
  &\times \exp\left[-\bigg(\frac{2}{L_{f}^{(\text{eff})}(n)}+\frac{1}{L_{b}^{(\text{eff})}(n)}\bigg)L l^2\right]
\notag\\
  &\times \left\{1 - \exp\left[-\bigg(
  iq+ \frac{2l+1/2}{L_{f}^{(\text{eff})}(n)}
  +\frac{l+1/2}{L_{b}^{(\text{eff})}(n)}\bigg)
  \left(L/2-z'\right) \right]
  \right\}\ .
\end{align}

All the remaining terms in matrix elements \eqref{tild_G_ij(z,z')} are averaged in a similar way, thus allowing,  after inverse Fourier transformation over $q$, for the full trial Green function. However, the exact expression for this function is rather cumbersome and not all that necessary for our purposes. To find the reflection coefficient it is more convenient to employ Eqs.~\eqref{G-t} and \eqref{t_n}, whereas the trial Green function expression valid for the entire interval $\mathbb{L}$ is required for the purposes of mode-mixing operator norm estimation only. For these purposes it would suffice to keep only those terms in Eqs.~\eqref{tild_G_ij(z,z')} that qualitatively represent the trial Green function dependence on the entire set of parameters, without taking care of quantitative validity of such an approximate representation.

From the detailed analysis it appears that for the order-of-magnitude norm estimation it would be reasonable to retain only the first terms of Eqs.~\eqref{tild_G_11(z,z')} and \eqref{tild_G_22(z,z')} among all formulas Eq.~\eqref{tild_G_ij(z,z')}, thereby forming the estimate-oriented \emph{model} trial Green function. Upon averaging, this function becomes
\begin{align}\label{Gn(z,z')-Av-3}
  \AV{G_n^{(V)}(z,z')}_{(mod)}= & \frac{\Phi_n}{2i\widetilde{k}_n}
  \exp\bigg\{\left[i \widetilde{k}_n - 1/2L_{\text{ext}}(n)\right]|z-z'|\bigg\}\ ,
\end{align}
where
\begin{align}\label{Phi_n-int}
  \Phi_n=\frac{1}{\sqrt{\pi}}\int\limits_{-\infty}^{\infty}dt
  \mathrm{e}^{-t^2}
  \left\{
  1-\left(\,\EuScript{R}_n \mathrm{e}^{i \widetilde{k}_n L}\right)^{2}
  \exp\left[\vphantom{\sqrt{\Bigg(\frac{1}{L_{f}^{(\text{eff})}(n)}\Bigg)}}\right.\right.
  & -\Bigg(\frac{1}{L_{f}^{(\text{eff})}(n)}+\frac{1}{L_{\text{ext}}(n)}\Bigg)|z-z'|
  \notag\\*
  & \left. \left. +2it\sqrt{L\Bigg(\frac{1}{L_{f}^{(\text{eff})}(n)}+\frac{1}{L_{\text{ext}}(n)}\Bigg)}\,\right]\right\}^{-1}\ ,
\end{align}
with notation
\begin{equation}\label{Extict_length(n)}
  \frac{1}{L_{\text{ext}}(n)}=\frac{1}{L_{f}^{(\text{eff})}(n)}+\frac{1}{L_{b}^{(\text{eff})}(n)}
\end{equation}
for the full inverse extinction length, is the factor containing the totality of parameters related to the particular mode scattering both due to the disorder (scattering lengths $L_{f,b}^{(\text{eff})}(n)$) and due to the mode mismatch at the ends of the corrugated section (coefficient $\EuScript{R}_n$). In the limiting cases of ballistic (with respect to the disorder-induced scattering) and Anderson-localized transport regimes factor \eqref{Phi_n-int} is computed to the following asymptotic values,
\begin{equation}\label{Phi_n-asympt}
  \Phi_n=
 \begin{cases}
   \mathcal{Q}_n, & \text{if\quad $L_{\text{ext}}(n)\gg L$}
   \\
   1\quad , & \text{if\quad $L_{\text{ext}}(n)\ll L$}
 \end{cases}
 \quad .
\end{equation}
%

\section{Evaluation of correlator Eq.~(\ref{Main_correlator})}
\label{R_l(n)-calc}

By differentiating function $R_l(z|n)$ over coordinate $z$ we get the equation
\begin{equation}\label{Main_correlator_z_eq1}
  \frac{d R_l(z|n)}{d z} = l  \left<\widetilde{\zeta}_-^*(z|n)\left[
  \widetilde{\Gamma}^{l}_-(z|n) \widetilde{\Gamma}^{* l-1}_-(z|n)+
  \widetilde{\Gamma}^{l+1}_-(z|n) \widetilde{\Gamma}^{* l}_-(z|n)\right]\right> + \ \text{c.c.}
\end{equation}
With Furutsu-Novikov formula and equalities~\eqref{VarDerivatives}, this equation can be converted to the following difference-differential form
\begin{equation}\label{Main_correlator_z_eq}
  \frac{d R_l(z|n)}{d z} =  \frac{l^2}{L_{b}^{(\text{eff})}(n)}
  \Big[R_{l+1}(z|n) - 2 R_l(z|n) + R_{l-1}(z|n) \Big]\ .
\end{equation}
The solution to Eq.~\eqref{Main_correlator_z_eq} must satisfy the conditions
\begin{subequations}\label{R_l(n)-conds}
  \begin{align}\label{R_l(n)-cond_1}
   & R_0(z|n) = 1\ ,\\
 \label{R_l(n)-cond_2}
   & R_l(-L/2|n) =  \EuScript{R}_n^{2 l}\ ,
  \end{align}
\end{subequations}
of which the first one stems from definition~\eqref{Main_correlator} whereas the second equality results from the ``initial'' condition for function $\widetilde{\Gamma}_-(z|n)$. Besides, from the flux conservation law taken in the form Eq.~\eqref{Pi->Gamma} one can notice that ${\big|\widetilde{\Gamma}_-(z|n)\big|^2 \leqslant 1}$. Therefore, as the third condition for function $R_l(z|n)$ we will make use of its convergence to zero with an infinite growth of index $l$.

The solution to equation~\eqref{Main_correlator_z_eq} may be sought in the integral form
\begin{equation}\label{R_l(z)->P_L(rho,z)}
  R_l(z|n) =  \int\limits_0^1 d \rho \widetilde{P}_L(\rho,z|n) \rho^l\ , \qquad l = 0,\,1,\,2,\ldots\ .
\end{equation}
Here, function $\widetilde{P}_L(\rho,z|n)$ can be regarded as the distribution function for random quantity $\big|\widetilde{\Gamma}_-(z|n)\big|^2$, provided that the requirement of normalization is met
\begin{equation}\label{P_L(rho,z)_norm}
  \int\limits_0^1 d \rho \widetilde{P}_L(\rho,z|n) = 1\ .
\end{equation}
This function satisfies the Fokker--Plank equation
\begin{subequations}\label{P_L(rho,z)}
\begin{equation}\label{P_L(rho,z)_eq}
  L_{b}^{(\text{eff})}(n) \frac{\partial \widetilde{P}_L(\rho,z|n)}{\partial z} =
  \frac{\partial}{\partial \rho} \rho \frac{\partial}{\partial \rho} (1 - \rho)^2 \widetilde{P}_L(\rho,z|n)
\end{equation}
and the condition
\begin{equation}\label{P_L(rho,z)_initial_cond}
  \widetilde{P}_L(\rho,-L/2|n) =  \delta\big(\rho-\EuScript{R}_n^2\big)\ .
\end{equation}
\end{subequations}

By going over to new variable $u$ through equality $\rho = (u-1)/(u+1)$ and representing function $R_l(z|n)$ as the integral
\begin{equation}\label{R_l(z)->P_L(u,z)}
  R_l(z|n) =  \int\limits_1^\infty d u P_L(u,z|n) \left(\frac{u-1}{u+1}\right)^l\ ,
\end{equation}
for function  $P_L(u,z|n)$, which is also normalized to unity, the following equation is derived
\begin{subequations}\label{P_L(u,z)_eq+init}
\begin{equation}\label{P_L(u,z)_eq}
  L_{b}^{(\text{eff})}(n) \frac{\partial P_L(u,z|n)}{\partial z} =
  \frac{\partial}{\partial u} (u^2 - 1) \frac{\partial P_L(u,z|n)}{\partial u}
\end{equation}
with the condition
\begin{equation}\label{P_L(u,z)_initial_cond}
  P_L(u,-L/2|n) =  \delta\left(u - \frac{1 + \EuScript{R}_n^2}{1 - \EuScript{R}_n^2}\right)\ .
\end{equation}
\end{subequations}
One can solve this equation through Mehler--Fock transformation,
\begin{align}
  P_L(u,z|n) =  \int\limits_0^\infty  d\lambda\,  \lambda \tanh(\pi \lambda)
  P_{-\frac{1}{2} + i \lambda}&(u)
  P_{-\frac{1}{2} + i \lambda}\left(\frac{1 + \EuScript{R}_n^2}{1 - \EuScript{R}_n^2}\right)
  \exp\left[- \left(\frac{1}{4}+\lambda^2\right) \frac{L + 2 z}{2 L_{b}^{(\text{eff})}(n)}\right]\ ,\label{P_L(u,z)}
\end{align}
and therefore obtain the analysis-friendly expression for requisite function~\eqref{Main_correlator}.

\section{The estimate of the mode-mixing operator norm}
\label{R_norm-estim}

Based upon standard definition of the operator norm \cite{KolmFom68,Kato66}, we have to evaluate the quantity
\begin{equation}\label{Norm-def}
  \Av{\|\hat{\mathsf{R}}\|^2}=\sup_{\psi}
  \frac{\Av{\big(\hat{\mathsf{R}}\psi,\hat{\mathsf{R}}\psi\big)}}{\big(\psi,\psi\big)}\ ,
\end{equation}
where function $\psi(z)$ belongs to the functional space where operator $\hat{\mathsf{R}}$ takes effect. As was shown in Section~\ref{ModeSep}, this is the coordinate-mode space ${\mathsf{M}}$, where the inner product includes both the integration over coordinate interval $\mathbb{L}$ and the summation over mode indices. The averaging in Eq.~\eqref{Norm-def} takes account of scattering potentials' random nature.

To avoid tedious calculations, we consider the action of operator $\hat{\mathsf{R}}=\hat{\mathcal{G}}^{(V)}\hat{\mathcal{U}}$ on the trial vector-valued function with $n$'th component only being different from zero and having the form $\psi_{q_n}(z)=\exp(iq_nz)\theta(L/2-|z|)$. At the last stage, the result should be maximized by the component number. Besides, in view of potential~\eqref{Unm} peculiar differential structure, it is more convenient to calculate not directly the operator $\hat{\mathsf{R}}$ norm but, rather, the norm of its transposed counterpart, whose matrix elements are given as
\begin{equation}\label{Transposed_R}
  \big(\hat{\mathsf{R}}^\mathrm{T}\big)_{ik}(z,z')=
  \hat{U}_{ki}(z)G_{k}^{(V)}(z',z)\ .
\end{equation}
With regard for the above remarks we need to make the estimate of the integral
\begin{align}\label{Norm-def2}
  \Av{\|\hat{\mathsf{R}}^\mathrm{T}\|^2}= & \sup_{\genfrac{}{}{0pt}{}{n\in\mathbb{Z}}{q_n\in\mathbb{R}}}\frac{1}{L}
  \int_{\mathbb{L}}dz\sum_{i\neq n}
  \iint_{\mathbb{L}}dz_1dz_2
  \Big<\hat{U}_{ni}^*(z)\hat{U}_{ni}(z){G_{n}^{(V)}}^*(z_1,z)G_{n}^{(V)}(z_2,z)\Big>
  \mathrm{e}^{-iq_n (z_1-z_2)}
  \notag\\
  \approx & \sup_{\genfrac{}{}{0pt}{}{n\in\mathbb{Z}}{q_n\in\mathbb{R}}}\frac{1}{L}
  \int_{\mathbb{L}}dz\sum_{i\neq n}
  \iint_{\mathbb{L}}dz_1dz_2
  \left<\big|\hat{U}_{ni}(z)\big|^2\right>\AV{{G_{n}^{(V)}}^*(z_1,z)G_{n}^{(V)}(z_2,z)}
  \mathrm{e}^{-iq_n (z_1-z_2)}\ ,
\end{align}
where the last equality is the approximate one since we neglect (with parametric accuracy!) the mean cross-products of intermode and intramode potentials.

In calculating the average square modulus of the potential in the rhs of Eq.~\eqref{Norm-def2} it is sufficient to restrict oneself to the second, linear in $w'(z)$, term in the rhs of Eq.~\eqref{Unm}. This statement is based on the following considerations. Let us split the potential \eqref{Unm} into its average value, $\Av{\hat{U}_{in}(z)}$, and the two fluctuating summands, one of which, $\Delta\hat{U}_{in}^{(sq)}(z)$, is related to square-in-$w'(z)$ term whereas the other, $\Delta\hat{U}_{in}^{(lin)}(z)$, is linear in the roughness slope. The average intermode potential is readily evaluated, and the result is as follows,
\begin{equation}\label{Av{Unm}}
  \Av{\hat{U}_{in}(z)}=\frac{8B_{in}D_{in}}{w_0^2}\Xi^2\ .
\end{equation}
The dispersion of square-gradient term in Eq.~\eqref{Unm} is of the same value. To calculate the linear term dispersion one should act by this term upon the trial function of the form $\exp(i\widetilde{k}_nz)$, whereupon the estimate follows
\begin{equation}\label{DeltaU(1)/DeltaU(2)}
  \sqrt{\frac{\Av{\big|\Delta\hat{U}_{in}^{(sq)}(z)\big|^2}}{\Av{\big|\Delta\hat{U}_{in}^{(lin)}(z)\big|^2}}}
  \sim \frac{\Xi}{w_0\sqrt{|\widetilde{k}_n|^2+1/r_c^2}}
  \sim\frac{\sigma}{w_0(1+kr_c)}\ll 1\ ,
\end{equation}
which enables us to keep the linear term only in Eq.~\eqref{Unm} when calculating the integrals on the right-hand side of~\eqref{Norm-def2}.

The trial Green functions entering into Eq.~\eqref{Norm-def2} are presented by expressions~\eqref{G-matrix} and \eqref{G_ij(z,z')}, whose exact form is excessively complicated for estimation purposes. In \ref{Mod_Trial_Green_func} it is shown that in view of configurational averaging, function \eqref{G-matrix} is qualitatively equivalent to more simple function
\begin{subequations}\label{Edge_res-interp}
\begin{align}\label{Edge_res-interp-1}
  \bm{\widetilde{G}}_n^{(V)}(z,z') & \simeq\frac{\mathfrak{A}_n}{2i\widetilde{k}_n}
  \exp\bigg[\left(i \widetilde{k}_n - \frac{1}{2L_{\text{loc}}(n)}\right)|z-z'|\bigg] \ ,\\
\intertext{where coefficient $\mathfrak{A}_n$ is equal to}
  \mathfrak{A}_n & =\mathcal{Q}_n\theta
  \big[L_{\text{loc}}(n)-L\big]+\theta\big[L-L_{\text{loc}}(n)\big]\ .
\label{Edge_res-interp-2}
\end{align}
\end{subequations}
So, we will use just this particular function when estimating expression~\eqref{Norm-def2}. As the length parameter $L_{loc}(n)$ in Eq.~\eqref{Edge_res-interp-1}, the quantity of the order of backscattering length ${L_{\text{loc}}(n)\sim L_{b}^{(\text{eff})}(n)}$ will be taken, where
\begin{equation}\label{L_b(eff)}
  \frac{1}{L_{b}^{(\text{eff})}(n)}= \frac{1}{L_{b}^{(s)}(n)}+\frac{1}{L_{b}^{(h)}(n)}\ ,
\end{equation}
thereby highlighting the well-known fact \cite{LifGredPast88} that, after averaging the binary combinations of complex conjugated Green functions for one-dimensional systems, the spatial attenuation of correlators subject to calculation is determined by localization length $L_{\text{loc}}(n)=4L_{b}^{(\text{eff})}(n)$.

By representing the function~\eqref{Edge_res-interp-1} as Fourier integral
\begin{equation}\label{G(V)-Furier}
  \bm{\widetilde{G}}_n^{(V)}(z,z')\simeq \mathfrak{A}_n\int\limits_{-\infty}^{\infty}\frac{d\varkappa}{2\pi}
  \frac{\mathrm{e^{i\varkappa(z-z')}}}{\widetilde{K}_n^2-\varkappa^2}\ ,
\end{equation}
where $\widetilde{K}_n=\widetilde{k}_n+i/2L_{\text{ext}}(n)$, we can recast Eq.~\eqref{Norm-def2} in the form
\begin{align}\label{R_norm_def-3}
  \Av{\|\hat{\mathsf{R}}^\mathrm{T}\|^2}= &
  \frac{16}{w_0^2}\sup_{\genfrac{}{}{0pt}{}{n\in\mathbb{Z}}{q_n\in\mathbb{R}}}
  |\mathfrak{A}_n|^2\sum_{i\neq n}B_{ni}^2
  \iint\limits_{-\infty}^{\quad\infty}\frac{d\varkappa_1 d\varkappa_2}{(2\pi)^2}
  \frac{1}{\Big(\widetilde{K}_n^{2*}-\varkappa_1^2\Big)\Big(\widetilde{K}_n^{2}-\varkappa_2^2\Big)}
  \notag\\*
 & \times
  \int_{\mathbb{L}}\frac{dz}{L}\mathrm{e}^{i(\varkappa_1-\varkappa_2)z}
  \AV{\big[\xi''(z)+2i\varkappa_1\xi'(z)\big]\big[\xi''(z)-2i\varkappa_2\xi'(z)\big]}
  \notag\\*
 &\times
  \iint_{\mathbb{L}}dz_1dz_2\mathrm{e}^{-i(\varkappa_1+q_n)z_1+i(\varkappa_2+q_n)z_2}
 \notag\\
 =& 16\left(\frac{\sigma}{w_0}\right)^2\sup_{\genfrac{}{}{0pt}{}{n\in\mathbb{Z}}{q_n\in\mathbb{R}}}
  |\mathfrak{A}_n|^2\sum_{i\neq n}B_{ni}^2
  \iint\limits_{-\infty}^{\quad\infty}\frac{d\varkappa_1 d\varkappa_2}{(2\pi)^2}
  \frac{W^{(IV)}(0)-4\varkappa_1\varkappa_2W''(0)}
  {\Big(\widetilde{K}_n^{2*}-\varkappa_1^2\Big)\Big(\widetilde{K}_n^{2}-\varkappa_2^2\Big)}
 \notag\\*
  & \times\frac{1}{L}\Delta(\varkappa_1-\varkappa_2)\Delta(\varkappa_1+q_n)\Delta(\varkappa_2+q_n)
  \ .
\end{align}
Here the notation
\begin{equation}\label{Delta-prelimit}
  \Delta(\kappa)=
  \int_{\mathbb{L}}dz\exp(\pm i\kappa z)=
  \frac{\sin(\kappa L/2)}{\kappa/2}\xrightarrow[L\to\infty]{}2\pi\delta(\kappa)
\end{equation}
is used for the function that possesses a sharp maximum at the zero argument and whose characteristic width is of the order of $1/L$.

It can be easily seen that in the integrals over $\varkappa_{1,2}$ in Eq.~\eqref{R_norm_def-3} two types of sharp functions are available. On the one hand, there is the Fourier transform of trial Green function~\eqref{Edge_res-interp-1}, and on the other is the prelimit $\delta$-function~\eqref{Delta-prelimit}. The competition between these two sharp functions determines actually two feasible different regimes of the given waveguide mode transport. The first mechanism, which corresponds to inequality ${L\ll L_{\text{loc}}(n)}$ and hence to predominant sharpness of trial Green functions in Eq.~\eqref{R_norm_def-3}, will be referred to as the (quasi)-ballistic regime. The other regime, where the opposite inequality holds $L\gg L_{\text{loc}}(n)$, will be referred to as the ``localized'' regime. Consider these two regimes separately.

1) \emph{``Ballistic'' regime}, $L\ll L_{\text{loc}}(n)$.

In this limiting case the sharpest functions in the integrand of Eq.~\eqref{R_norm_def-3} are the Green function Fourier transforms, whose maxima are at $\varkappa_{1,2}=\widetilde{k}_n$. By carrying the $\Delta$-functions out of integrals at these points we obtain
\begin{subequations}\label{R_norm}
\begin{align}\label{R_norm-ball_1}
  \Av{\|\hat{\mathsf{R}}^\mathrm{T}\|^2}= &
  4\left(\frac{\sigma}{w_0}\right)^2\sup_{\genfrac{}{}{0pt}{}{n\in\mathbb{Z}}{q_n\in\mathbb{R}}}
  |\mathfrak{A}_n|^2\frac{\Delta^2(\widetilde{k}_n+q_n)}{\widetilde{k}_n^2}
  \Big[W^{(IV)}(0)-4\widetilde{k}_n^2W''(0)\Big]
  \sum_{\genfrac{}{}{0pt}{}{i\leqslant N_c(\Xi)}{(i\neq n)}}B_{ni}^2
  \notag\\
  \sim & \left(\frac{\sigma}{w_0}\right)^2 \sup_{\genfrac{}{}{0pt}{}{n\in\mathbb{Z}}{}}
  L^2|\mathfrak{A}_n|^2
  \frac{\Big[1+\big(\widetilde{k}_nr_c\big)^2\Big]}
  {r_c^2\big(\widetilde{k}_nr_c\big)^2}\sum_{\genfrac{}{}{0pt}{}{i\leqslant N_c(\Xi)}{(i\neq n)}}
  \frac{n^2i^2}{(n^2-i^2)^2}\ .
\end{align}
The summation over $i$ in Eq.~\eqref{R_norm-ball_1} is limited by extended modes only, since the evanescent modes attenuate over the distance of the order of mode wavelength and so contribute slightly to the extended mode scattering.

2) \emph{``Localized'' regime}, $L\gg L_{\text{loc}}(n)$.

In this case the $\Delta$-functions in Eq.~\eqref{R_norm_def-3} can be replaced with true $\delta$-functions, resulting~in
\begin{align}\label{R_norm-loc_1}
  \Av{\|\hat{\mathsf{R}}^\mathrm{T}\|^2}= &
  16\left(\frac{\sigma}{w_0}\right)^2\sup_{\genfrac{}{}{0pt}{}{n\in\mathbb{Z}}{q_n\in\mathbb{R}}}
  |\mathfrak{A}_n|^2\frac{W^{(IV)}(0)-4q_n^2W''(0)}{\Big|\widetilde{K}_n^{2}-q_n^2\Big|^2}
  \sum_{\genfrac{}{}{0pt}{}{i\leqslant N_c(\Xi)}{(i\neq n)}}B_{ni}^2
\notag\\
  \sim & \left(\frac{\sigma}{w_0}\right)^2 \sup_{\genfrac{}{}{0pt}{}{n\in\mathbb{Z}}{q_n\in\mathbb{R}}}
  L_{\text{loc}}^2(n)|\mathfrak{A}_n|^2
  \frac{\Big[1+\big(\widetilde{k}_nr_c\big)^2\Big]}
  {r_c^2\big(\widetilde{k}_nr_c\big)^2}\sum_{\genfrac{}{}{0pt}{}{i\leqslant N_c(\Xi)}{(i\neq n)}}\frac{n^2i^2}{(n^2-i^2)^2}\ .
\end{align}
\end{subequations}
The difference of this formula from Eq.~\eqref{R_norm-ball_1} consists in the dimensional factor standing in front of factor $|\mathfrak{A}_n|^2$, therefore both of the estimates~\eqref{R_norm} may be cast in the form of a single estimation formula
\begin{align}\label{R_norm-united}
  \Av{\|\hat{\mathsf{R}}^\mathrm{T}\|^2}\sim \left[1-\delta_{1,N_c(\Xi)}\right]\left(\frac{\sigma}{w_0}\right)^2
  \sup_{\genfrac{}{}{0pt}{}{n<N_c(\Xi)}{}}
  |\mathfrak{A}_n|^2 n^2
  \frac{\min\Big\{L^2,\big[2L_{\text{ext}}(n)\big]^2\Big\}}{r_c^2}
  \cdot\frac{1+\big(\widetilde{k}_nr_c\big)^2}{\big(\widetilde{k}_nr_c\big)^2}
  \ .
\end{align}
In the course of its derivation we have considered that only the terms with $i\sim n$ contribute effectively to the sums in Eqs.~\eqref{R_norm}, and that the number of essential terms is of the order of unity.



\bibliography{refs}

\end{document}